# Systematic design and realization of double-negative acoustic metamaterials by topology optimization

Hao-Wen Dong[1,2,†], Sheng-Dong Zhao[3,†], Peijun Wei[1], Li Cheng[2], Chuanzeng Zhang[4], Yue-Sheng Wang[5*]

[1]*Department of Applied Mechanics, University of Science and Technology Beijing, Beijing 100083, PR China*

[2]*Department of Mechanical Engineering, The Hong Kong Polytechnic University, Hong Kong, PR China*

[3]*School of Mathematics and Statistics, Qingdao University, Qingdao 266071, PR China*

[4]*Department of Civil Engineering, University of Siegen, D-57068 Siegen, Germany*

[5]*Department of Mechanics, School of Mechanical Engineering, Tianjin University, Tianjin 300350, PR China*

**Highlights:**
- Unified topology optimization framework is developed for designing double-negative acoustic metamaterials (AMMs).
- Representative resonance-cavity-based and space-coiling microstructures are explored.
- Broadband double negativity originating from novel multipolar LC (inductor-capacitor circuit) or Mie resonances can be induced by easily controlling optimization parameters.
- Desired broadband subwavelength imaging of topology-optimized AMMs is verified experimentally.

**Abstract:**
Acoustic metamaterials (AMMs) with negative parameters enable novel ways of focusing and shaping wave fields at subwavelength scales. Double-negative AMMs offer the promising ability of superlensing for applications in ultrasonography, biomedical sensing and nondestructive evaluation. However, the systematic design and realization of broadband double-negative AMMs is stilling missing, which hinders their practical implementations. In this paper, under the simultaneous increasing or non-increasing mechanisms, we develop a unified topology optimization framework considering the different microstructure symmetries, minimal structural feature sizes and dispersion extents of effective parameters. Then we apply the optimization framework to furnish the heuristic resonance-cavity-based and space-coiling metamaterials with broadband double negativity. Meanwhile, we demonstrate the essences of double negativity derived from the novel artificial multipolar LC (inductor-capacitor circuit) and Mie resonances which can be induced by controlling mechanisms in optimization. Furthermore, abundant numerical simulations validate the double negativity, negative refraction, enhancements of evanescent waves and subwavelengh imaging for the optimized AMMs. Finally, we experimentally show the desired broadband subwavelengh imaging using the 3D-printed optimized space-coiling metamaterial. The present design methodology provides an ideal approach for constructing the constituent "atoms" of metamaterials according to any manual physical and structural requirements. In addition, the optimized broadband AMMs and superlens can truly lay the structural foundations of subwavelengh imaging technology.

---

[†] These authors contributed equally to this work.
[*] Corresponding author, yswang@tju.edu.cn

## 1. Introduction

Due to the fantastic wave characteristics, metamaterials (Pendry, 2000; Fang et al., 2006; Valentine et al., 2008; Han et al., 2014; Frenzel et al., 2017; Matlack et al., 2018) designed by engineering the subwavelength microstructures offer novel and surprising opportunities for manipulating and controlling wave propagation, revealing the broad application prospects in the fields of mechanics, materials, optics, electromagnetism, acoustics, and thermotics, etc. In general, conventional materials drive their wave motions from the properties of intrinsic atoms or molecules; metamaterials provide infinite possibilities for constructing the artificial "meta-atoms" (microstructures) with special geometry, physical features and spatial arrangements, thus bringing out much new functionality. Electromagnetic metamaterials and metadevices can achieve a range of exotic electromagnetic responses, including negative refractive index, zero refractive index, optical chirality, anisotropy and hyperbolicity. Inspired by the optical metamaterials, acoustic metamaterials (AMMs) (Fang et al., 2006; Liang et al., 2012; Liu et al., 2018), elastic metamaterials (Liu et al., 2000; Lai et al., 2011; Zhu et al., 2014; Dong et al., 2017, 2018), mechanical metamaterials (Frenzel et al., 2017) and even graphene metamaterials (Lee et al., 2012) have been developed in many ways. Like the other types of metamaterials, creating the suitable building blocks of microstructures is the most fundamental and pivotal point for AMMs which exhibit the diverse combinations of the effective constitutive parameters–the mass density $\rho_{eff}$ and bulk modulus $K_{eff}$. In the quadrants of AMMs, the reported representative cases are single negativity ($\rho_{eff}<0$, $K_{eff}>0$; $\rho_{eff}>0$, $K_{eff}<0$), double negativity ($\rho_{eff}<0$, $K_{eff}<0$), double positivity ($\rho_{eff}>0$, $K_{eff}>0$) near-zero mass density ($\rho_{eff}\approx0$), and even double-zero index ($\rho_{eff}\approx0$, $1/K_{eff}\approx0$). Benefitting from the exotic effective properties, AMMs possess the great potential applications in low-frequency isolation (Kumar et al., 2018), space sound field modulating (Ma et al., 2018), energy harvesting (Qi et al., 2016), perfect absorption (Mei et al., 2012), negative refraction (Liang et al., 2012), cloaking (Zhang et al., 2011) and nonreciprocal acoustic devices (Popa et al., 2014), and thus attracting the widespread and continuous attention during the past two decades. In acoustics, one of the most promising ability of AMMs is the subwavelength superlensing, leading to high-resolution ultrasonic imaging for medicine and industry (Zhang et al., 2008; Kaina et al, 2015). Although the anamorphic effective refractive index (Zigoneanu et al., 2011) or phase difference (Li et al., 2014) can enable the gradient metamaterials to focus waves in a focal plane, the conspicuous shortcoming is that their imaging resolutions cannot essentially break the diffraction limit. Alternatively, several strategies using the microstructures (Li et al, 2009; Zhu et al., 2011; Park et al., 2015; Lanoy et al., 2015) with different features of effective parameters can collect and exploit the evanescent wave field for the subwavelength details. One prominent technique for subwavelength imaging is the double-negative superlens (Park et al., 2015). It can cause the negative refraction, then bring the diverging waves to reconvene and amplify the evanescent waves in the near field. Another approach relies on the anisotropic metamaterials (Li et al, 2009; Zhu et al., 2011) which can convert the coupling of near-fields emitted by subwablength objects into propagating waves. In addition to the above strategies, time-reversal technique can also control and focus the subwavelength waves. By virtue of the Helmholtz resonators, the temporal response is recorded, flipped in time and radiated back, achieving the subwavelength focusing (Lanoy et al., 2015). Furthermore, the recent research indicates that, for the metamaterials having relatively high index within a slow medium, the excitation of guided acoustic modes can transmit only subwavelength information to generate the subwavelength edge-based imaging (Molerón et al., 2015). In consideration of the operating bandwidth, in particular, the double-negative and hyperbolic metamaterials are more suitable for the broadband acoustic subwavelength focusing and imaging applications. In most cases, the hyperbolic metamaterials have quite high demands for the structure design. Only the three-dimensional membrane-type metamaterials (Shen et al., 2015) or layers of perforated plates (Christensen et al., 2012) can produce the extremely anisotropic dispersion relations at the subwavelength scale. Nevertheless, the double-negative metamaterials afford three choices including the coupled filter-element (Lee et al., 2010), coupled-membrane (Yang et al., 2013) and space-coiling structures (Liang et al., 2012). Therefore, it is necessary to delve into constructing the double negativity and the corresponding subwavelength superlensing.

In principle, the acoustic double negativity implies the several building blocks or specific elements supporting multiple overlapping resonances. The first approach is combining two resonating structures, membranes and Helmholtz resonators (Lee et al., 2010), to guarantee that their symmetric eigenmodes occur in the same dispersive frequency range, obtaining the double negativity from 240 to 450 Hz based on a periodic array of interspaced membranes and side holes. Alternatively, coupled-membrane resonators can also lead to double negativity with the monopolar and dipolar eigenmodes in the range of 520–830 Hz (Yang et al., 2013). Another strategy relying on the ultra-slow Mie resonators consisting of macroporous microbeads (Brunet et al., 2015) can also acquire the negative acoustic index. Moreover, by coiling up space, the labyrinthine structures composed of hard solid plates inserted into the background fluid can cause the large phase delay and form the band folding at the low frequency ranges (Liang et al., 2012), and thus exhibit a frequency dispersive spectrum of a large refractive index not found in nature and double negativity without the traditional resonant elements. While the aforementioned methods preliminarily realized double negativity, the following problems and challenges need to be solved to capture the broadband double negativity for the subwavelength imaging. Firstly, from the perspective of microstructure design, the primary double-negative AMMs mainly depend on the membrane or space-coiling structures. Compared with the membrane-type metamaterials, the space-coiling structures have attracted more considerable attention for its strong control over the effective parameters and easy implementation. Because of the difficulties for constructing the pertinent zigzag path, however, much research only paid attention to the flexible phase manipulation (Xie et al., 2014) instead of the broadband double negativity within the spectrum of interest. Meanwhile, except the space-coiling metamaterials, designing more forms of double-negative solid-air AMMs becomes a pressing issue to get more choices for practical imaging applications. Secondly, in regard of double-negative mechanism, existing AMMs mainly utilize the LC (inductor-capacitor circuit) resonance (Lee et al., 2010) induced from the different resonant elements or the Mie resonance (Brunet et al., 2015) produced by the particles with high refractive index relative to the background medium. But there are few works about the LC-resonance double negativity based on the artificial structures without membranes. Mie-resonance double negativity in a broadband range also challenges the microstructure design. Thirdly, with respect to the double-negative bandwidth, resonance-based AMMs usually have frequency dispersions and narrow bands. Hence, the obvious direction of double-negative AMMs should be broadening the frequency range for solidifying their roles in diverse applications (Cummer et al., 2016). But in general, the above three issues are collectively limited by the manual and empirical design strategy.

Encouraged by the burgeoning 3D printing, topology optimization have successively accomplished the metamaterials design and then reached the desired performances (Otomori et al., 2012, 2017; Sanchis et al., 2013; Dong et al., 2017, 2018; Christiansen et al., 2016; Yang et al., 2018; Wang et al., 2018). Since AMMs have showed the unprecedented functionalities on wave manipulations, several topology optimization studies of AMMs are subsequently reported in recent years (Li et al., 2012; Lu et al.; 2013; Christiansen et al., 2016). However, these works focus on the optimization of propagation responses (Christiansen et al., 2016) or positive wave parameters (Li et al., 2012). Furthermore, the topology-optimized AMMs only completed the expected narrow-band single negativity (Lu et al.; 2013), anisotropic dispersion relation (Christiansen et al., 2016) and negative refraction (Christiansen et al., 2016), lacking of the demonstration of subwavelength imaging. So topology optimization of AMMs is just unfolding and becomes an immediate demand for more newfangled phenomena and acoustic devices. Up to now, the inverse design of double-negative AMMs for airborne sound is still missing, let alone the systematic optimization study. Moreover, unlike the elastic metamaterials (Dong et al., 2017, 2018), the negative effective parameters not only depend on the LC resonance, but also are possibly dominated by the Mie resonance.

Because of the doubtful mechanism for negativity, topology optimization of double-negative AMMs is full of challenges.

In this paper, to provide the comprehensive guidance on engineering the double-negative AMMs, we show for the first time that the broadband double-negative AMMs can be systematically designed by topology optimization. A unified topology optimization framework is constructed for obtaining the broadband double negativity as far as possible within the prescribed low frequency range. The proposed framework considers several typical structural and physical characteristics of the microstructure, including the unitcell's symmetry (i.e., the square, chiral and orthogonal symmetries), the minimal geometrical size (i.e., the minimal size of solid parts and width of air channels), the variation tendency and the frequency dispersion degree of effective parameters. Band structures and effective parameters retrieval demonstrate that the topology-optimized microstructures really exhibit the broadband double negativity. All metamaterials present here uncover two kinds of topology characteristics, i.e., the resonance-cavity-based and space-coiling layouts. Lots of eigenstate analysis reveal that the optimization methodology under simultaneous increasing tendencies of the effective mass density and bulk modulus can give rise to the LC-resonance double negativity; whereas the simultaneous non-increasing tendencies can evoke the Mie-resonance double negativity. Then, the subwavelength negative refraction and acoustic imaging are numerically demonstrated after validating the equi-frequency surfaces and enhanced transmission of the evanescent waves. Finally, we fabricate a 3D-printed topology-optimized space-coiling AMM to successfully perform the subwavelength imaging in an acoustic experiment.

## 2. Topology optimization methodology

Consider the square-latticed microstructures consisting of solid and air elements for the acoustic wave propagation, as show in Fig. 1. Based on the simulation model depicted in Fig. 1(a), the transmission and reflection coefficients of one microstructure can be calculated for the acoustic effective parameters retrieval, as long as the microstructure is symmetric along the direction of wave propagation. The dispersion relations of the microstructures can be characterized by the Floquet-Bloch theory. In topology optimization, the microstructure is divided into $N \times N$ pixels, in which air and solid are denoted by "0" and "1", respectively. This paper presumes that the microstructure has three representative types of symmetries during topology optimization, i.e., the square chiral and orthogonal symmetries, as illustrated in Fig. 1(c). Because of the symmetry assumptions, the design domain in topology optimization changes from the whole unitcell to a reduced space, see Fig. 1(c). Similarly, their corresponding irreducible Brillouin zones are displayed in Fig. 1(c).

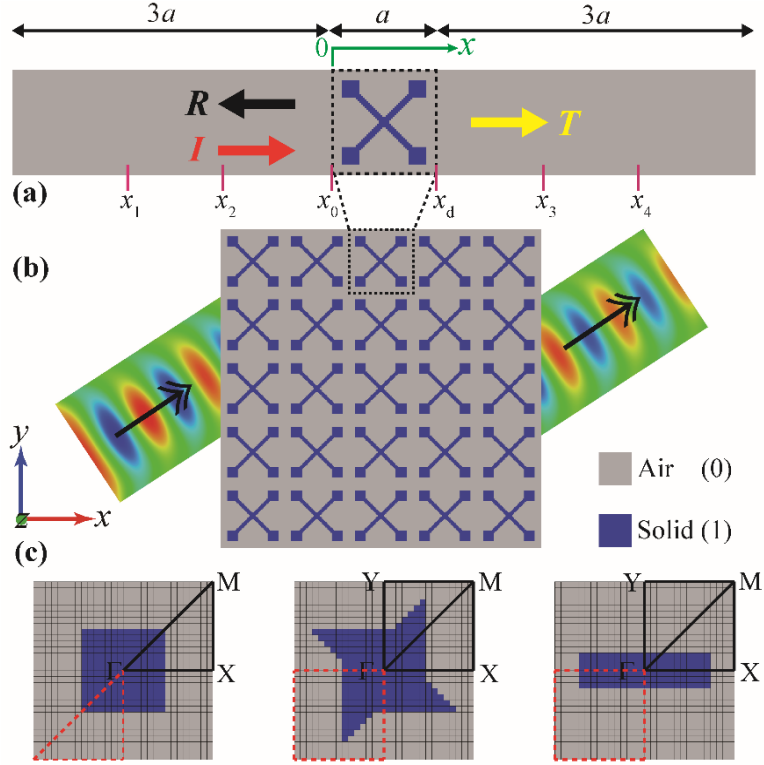

**Fig. 1. Schematic illustration of the square-latticed AMMs and considered microstructural symmetries. (a)** Wave propagation model for calculating the transmission and reflection coefficients of one microstructure. The simulation domain is terminated with the infinite elements at the left and right boundaries. The top and bottom edges are set as rigid walls. Four probes are introduced in the model for the scattering coefficients retrieval. **(b)** Double-negative AMMs with desired wave manipulation. **(c)** Three representative symmetries (left: square symmetry; middle: chiral symmetry; right: orthogonal symmetry) investigated in topology optimization. The dashed and solid lines show the corresponding reduced design domains and edges of the first reduced Brillouin zone, respectively.

## *2.1. Characterization of AMMs*

To formulate the wave equations for the present high contrast (solid/air) wave problems shown in Fig. 1, the elastic-acoustic interactions can be ignored for simplicity. Because the solid can be regarded as perfectly hard, that is, the wave propagation is principally predominant in the background air (Goffaux et al., 2001; Christiansen et al., 2016), it is a good approximation to take the solid as a fluid with very high stiffness and specific mass (Goffaux et al., 2001). Therefore, we only consider the traditional acoustic wave equation:

$$\nabla \cdot \left[ \rho^{-1}(\mathbf{r}) \nabla p(\mathbf{r}) \right] + \omega^2 \lambda^{-1}(\mathbf{r}) p(\mathbf{r}) = 0, \tag{1}$$

where $p$ is the acoustic pressure; and $\lambda$ for acoustic case equals to the bulk modulus $K$. According to the Floquet-Bloch theory, the acoustic pressure can be written as $p(\mathbf{r})=e^{i\mathbf{k}\cdot\mathbf{r}}p_{\mathbf{k}}(\mathbf{r})$, where $\mathbf{k}$ and $p_{\mathbf{k}}(\mathbf{r})$ are the Bloch wave vector of the first Brillouin zone and periodic function of $\mathbf{r}$, respectively. Considering the wave equation and boundary conditions, we can calculate the dispersion relation ($\omega$–$\mathbf{k}$) by using the ABAQUS/Standard solver Lanczos. Then the eigen-modes, phase velocity, group velocity and equi-frequency surfaces can be obtained from the band structures.

Under the long-wavelength assumption, a prominent trait of metamaterials is the feasibility of describing the artificial microstructure by dynamic effective medium theory (EMT). For characterizing the AMMs, the primary and commonly used approach is extracting the acoustic effective properties from reflection and transmission coefficients (Fokin et al., 2007). The essence of this inverse technique is reproducing the far-field scattering properties in an average sense through a uniform medium. To retrieve the effective constitutive parameters, the simulation model is regarded as a two-port network, shown in Fig. 1(a), in which an incident plane wave $I$ propagates normally to the microstructure, with the reflection ($R$) and transmission ($T$) coefficients, respectively. Two probes are placed at $x_1$ and $x_2$ in the front and the other two at $x_3$ and $x_4$ in the back of the microstructure. After acquiring the total pressure of four probes, the four-microphone method (Song et al., 2000) is adopted to calculate the scattering matrix **S** of the microstructure, thus getting the effective refractive index and impedance. Finally, the effective mass density and bulk modulus can be further determined. For the details, we refer to Ref. Fokin et al. (2007). It is noted that the above four-microphone method is very easy to be implemented through the acoustic experiment.

In our simulation, the complex sound pressures at the four probes $x_1$, $x_2$, $x_3$ and $x_4$ are respectively expressed as

$$P_1 = \left(Ae^{ik_0x_1} + Be^{-ik_0x_1}\right)e^{-i\omega t},$$
$$P_2 = \left(Ae^{ik_0x_2} + Be^{-ik_0x_2}\right)e^{-i\omega t},$$
$$P_3 = \left(Ce^{ik_0x_3} + De^{-ik_0x_3}\right)e^{-i\omega t},$$
$$P_4 = \left(Ce^{ik_0x_4} + De^{-ik_0x_4}\right)e^{-i\omega t},$$
(2)

where $A$, $B$, $C$ and $D$ denote the complex amplitudes of the positive- and negative-going plane waves; $k_0$ is the wave number in the background medium (air); $\omega$ is the circular frequency; $e^{-i\omega t}$ represents the common time-harmonic factor which is omitted throughout the paper for the sake of brevity; $x_1$, $x_2$, $x_3$ and $x_4$ indicate the corresponding distances of four probes relative to respective reference planes in the two ports. Since the pressures $P_1$, $P_2$, $P_3$ and $P_4$ can be directly obtained by the numerical simulation, the four coefficients $A$, $B$, $C$ and $D$ can be derived from Eq. (2) as

$$A = -i\frac{P_1 e^{-ikx_2} - P_2 e^{-ikx_1}}{2\sin(k_0 x_1 - k_0 x_2)},$$
$$B = -i\frac{P_2 e^{ikx_1} - P_1 e^{ikx_2}}{2\sin(k_0 x_1 - k_0 x_2)},$$
$$C = i\frac{P_3 e^{-ikx_4} - P_4 e^{-ikx_3}}{2\sin(k_0 x_4 - k_0 x_3)},$$
$$D = i\frac{P_4 e^{ikx_3} - P_3 e^{ikx_4}}{2\sin(k_0 x_4 - k_0 x_3)}.$$
(3)

The second order matrix relating the acoustic particle velocity and sound pressure on two faces of the microstructure in the simulation model shown in Fig. 1(a) is defined as the transfer matrix which is denoted as **T** with the elements $T_{ij}$ ($i, j$=1, 2). In view of the considered symmetries in Fig. 1(c), the effective two-port system (Song et al., 2000) is essentially reciprocal. In other words, **T** should satisfy

$T_{11}=T_{22}$, (4)

$T_{11}T_{22}−T_{12}T_{21}=1$, (5)

In addition, the simulation model has the following boundary conditions

$p_0=A+B$, (6)

$p_d=C+D$, (7)

$v_0=(A−B)/Z_0$, (8)

$v_d=(D−C)/Z_0$, (9)

where $p_0$ and $p_d$ are the pressures at locations $x_0$ and $x_d$ in Fig. 1(a), respectively; $v_0$ and $v_d$ express the particle velocities at locations $x_0$ and $x_d$ in Fig. 1(a), respectively; $Z_0$ is the impedance of the background medium. Then the transfer matrix of the effective two-port network can be written as

$$\mathbf{T} = \begin{bmatrix} T_{11} & T_{12} \\ T_{21} & T_{22} \end{bmatrix} = \begin{bmatrix} \dfrac{p_d v_d + p_0 v_0}{p_0 v_d + p_d v_0} & \dfrac{p_0^2 - p_d^2}{p_0 v_d + p_d v_0} \\ \dfrac{v_0^2 - v_d^2}{p_0 v_d + p_d v_0} & \dfrac{p_d v_d + p_0 v_0}{p_0 v_d + p_d v_0} \end{bmatrix}. \tag{10}$$

Based on the transformation relation between the scattering and transfer matrices (Song et al., 2000), the scattering matrix can be obtained as

$$\mathbf{S} = \begin{bmatrix} S_{11} & S_{12} \\ S_{21} & S_{22} \end{bmatrix} = \begin{bmatrix} \dfrac{T_{11} + T_{12}/Z_0 - T_{21}Z_0 - T_{22}}{T_{11} + T_{12}/Z_0 + T_{21}Z_0 + T_{22}} & \dfrac{2(-T_{21}T_{12} + T_{11}T_{22})}{T_{11} + T_{12}/Z_0 + T_{21}Z_0 + T_{22}} \\ \dfrac{2}{T_{11} + T_{12}/Z_0 + T_{21}Z_0 + T_{22}} & \dfrac{-T_{11} + T_{12}/Z_0 - T_{21}Z_0 + T_{22}}{T_{11} + T_{12}/Z_0 + T_{21}Z_0 + T_{22}} \end{bmatrix}, \tag{11}$$

where $S_{11}$ and $S_{21}$ are the reflection $R$ and transmission coefficient $T$, respectively. If the size of the microstructure is much smaller than the operating wavelength of the background medium (i.e., $\lambda_{bk} \geq 5a$), the composite microstructure can be regarded as the homogeneous medium (Popa et al., 2009; Xie et al., 2013), thus the effective impedance $Z_{eff}$ and effective refractive index $n_{eff}$ can be retrieved using the inverse technique (Fokin et al., 2007) in the form

$$Z_{eff} = \frac{\eta}{1 - 2R + R^2 - T^2}, \tag{12}$$

$$n_{eff} = \frac{-i \ln \xi + 2\pi m}{k_0 a}, \tag{13}$$

where $m$ represents the branch number of function $\cos^{-1}[(1-R^2+T^2)/2T]$; $\eta$ and $\xi$ are defined by

$$\eta = \mp \sqrt{(R^2 - T^2 - 1)^2 - 4T^2}, \tag{14}$$

$$\xi = \frac{1 - R^2 + T^2 + \eta}{2T}. \tag{15}$$

For the passive metamaterials, the physically meaningful natural feature is that the sign of $\eta$ should be chosen such that Re($Z_{eff}$) is positive. The calculation of $n_{eff}$ shown in Eq. (13) is highly depended on the value of $m$. For the thick metmamaterials, $m$ should be carefully selected and usually has finite values as integers. For the sake of simplicity, the metamaterial can be constructed with the minimal thickness whose size is much smaller than the wavelength so that $m=0$ can be guaranteed.

With the above determination of $Z_{eff}$ and $n_{eff}$, the effective mass density $\rho_{eff}$ and bulk modulus $K_{eff}$ are computed through

$$\rho_{eff} = \rho_0 Z_{eff} n_{eff}, \tag{16}$$

$$K_{eff} = \rho_0 c_0^2 Z_{eff} / n_{eff}, \tag{17}$$

where $\rho_0$ and $c_0$ are the mass density and acoustic velocity of the background medium, respectively. Meanwhile, the effective phase change $\Delta\phi_{eff}$ across the matamaterial layer can be achieved by $\Delta\phi_{eff} = \omega a \rho_{eff}/Z_{eff}$.

*2.2. Design problem formulation*

To realize a broadband double-negative AMM without membrane units, we need to construct a microstructure for resolving two emblematic challenges: (1) different resonance symmetries, including monopole, dipole and even quadrupole, have to be exploited through one microstructure; and (2) $\rho_{\text{eff}}$ and $K_{\text{eff}}$ should have the same dispersion property as the frequency increases. Fortunately, as a systematic mathematical method, topology optimization involves the optimization of material layout in the huge design space $2^{N \times N}$, providing the infinite possibilities for the occurrence of multiple resonances based on a brand-new topology. In general, the microstructure only holds single negativity (Cheng et al., 2015; Graciá-Salgado et al., 2013) or very narrow-band double negativity (Liang et al., 2012) if the variation tendencies of $\rho_{\text{eff}}$ and $K_{\text{eff}}$ are inconsistent. Consequently, it is essential to control their holistic properties and then guarantee the coincident performance.

From the previous studies on AMMs (e.g. Liang et al., 2012; Cheng et al., 2015), we find that the metamaterial usually possesses a relatively large $n_{\text{eff}}$ before generating the negative properties, whether the metamaterial has single negativity (Cheng et al., 2015) or narrow-band double negativity (Liang et al., 2012). More specially, the labyrinth microstructure has been demonstrated to be capable of achieving a long path length which is equivalent to a large $n_{\text{eff}}$ (Liang et al., 2012). As a result, the metamaterial can realize the band folding in the low-frequency range, thus obtaining the double negativity. For the other types of AMMs (Cheng et al., 2015; Graciá-Salgado et al., 2013) with single negativity, the negative $\rho_{\text{eff}}$ or $K_{\text{eff}}$ requires a resonance to generate an infinite effective value. That is, $\rho_{\text{eff}}$ will increase from a positive value to the negative infinity and then derive the negative values. However, $K_{\text{eff}}$ will decrease from a positive value to the infinity. Hence, in pursuing double negativity, $\rho_{\text{eff}}$ and $K_{\text{eff}}$ generally have the noticeable increasing and deceasing trends, respectively. In such situation, the relatively large $\rho_{\text{eff}}$ and relative small $K_{\text{eff}}$ will allow the microstructure to possess a large $n_{\text{eff}}$ over the whole concerned spectrum. Although $n_{\text{eff}}$ will show the obvious dispersion property, its value can keep the quasi-static feature in the ultra-low frequency range. Therefore, regardless of the resonance mechanism and structure topology, the common prerequisite condition for double negativity is getting the relatively large $n_{\text{eff}}$ in the ultra-low frequency range. When large $n_{\text{eff}}$ induces the suitable resonances, it is essential to avoid the single negative $\rho_{\text{eff}}$ or $K_{\text{eff}}$ by adjusting and controlling the dispersion extent of $\rho_{\text{eff}}$ and $K_{\text{eff}}$. In other words, the optimization model must properly punish the degree of variation for $\rho_{\text{eff}}$ and $K_{\text{eff}}$ over the whole concerned spectrum and improve $n_{\text{eff}}$ at the same time. Finally, for the aim of the acoustic broadband double negativity along the $x$ direction within the target frequency ranges, the consolidated optimization formulation considering the prescribed physical mechanisms of effective constitutive parameters and special structural feature sizes is proposed as

$$\text{For: } \omega \in [\omega_{\min}, \omega_{\max}] \tag{18}$$

$$\text{Maximize: } OF = N_{\text{n}} + \frac{n_{\text{eff}}^{x+,1} \times n_{\text{eff}}^{x+,1}}{M} - \frac{\alpha}{M} \times \max_{\substack{i=1,2\cdots M' \\ j=1,2\cdots N'}} \left[ \frac{\max(\rho_{\text{eff}}^{x+,i})}{\min(\rho_{\text{eff}}^{x+,i})}, \frac{\max(K_{\text{eff}}^{+,j})}{\min(K_{\text{eff}}^{+,j})} \right], \tag{19}$$

$$\text{Subject to: } \rho_i = 0 \text{ or } 1 \ (i = N \times N), \tag{20}$$

$$CD_{\text{air}} = 1, \tag{21}$$

$$\min_{\Sigma}(w_{\text{a}}) \geq w_{\text{a}}^*, \tag{22}$$

$$\min_{\Sigma}(w_{\text{s}}) \geq w_{\text{s}}^*, \tag{23}$$

$$\beta = \begin{cases} \beta_1 = \min_{i=1,2\cdots M'}\left[\rho_{\text{eff}}^{x+,2} - \rho_{\text{eff}}^{x+,1},\cdots,\rho_{\text{eff}}^{x+,i} - \rho_{\text{eff}}^{x+,i-1}\right] > 0 \\ \beta_2 = \min_{j=1,2\cdots N'}\left[K_{\text{eff}}^{+,2} - K_{\text{eff}}^{+,1},\cdots,K_{\text{eff}}^{+,j} - K_{\text{eff}}^{+,j-1}\right] > 0 \end{cases} \text{(case 1),}$$

$$\beta = \begin{cases} \beta_1 = \min_{i=1,2\cdots M'}\left[\rho_{\text{eff}}^{x+,2} - \rho_{\text{eff}}^{x+,1},\cdots,\rho_{\text{eff}}^{x+,i} - \rho_{\text{eff}}^{x+,i-1}\right] \leq 0 \\ \beta_2 = \min_{j=1,2\cdots N'}\left[K_{\text{eff}}^{+,2} - K_{\text{eff}}^{+,1},\cdots,K_{\text{eff}}^{+,j} - K_{\text{eff}}^{+,j-1}\right] \leq 0 \end{cases} \text{(case 2),}$$

(24)

where $\omega_{\min}$ and $\omega_{\max}$ are the upper and lower bounds of the target frequency range which is divided by $M$ sampling frequency points; $OF$ denotes the objective function; $N_n$ is the number of sampling frequency points where the double negativity is implemented; $n_{\text{eff}}^{x+}$, $\rho_{\text{eff}}^{x+}$ and $K_{\text{eff}}^{+}$ are the arrays of positive $n_{\text{eff}}^{x}$, positive $\rho_{\text{eff}}^{x}$ and positive $K_{\text{eff}}$, respectively; $M'$ and $N'$ are the numbers of the elements belong to the arrays of $\rho_{\text{eff}}^{x+}$ and $K_{\text{eff}}^{+}$, respectively; $n_{\text{eff}}^{x+,1}$ means the first positive effective refractive index of an array of $n_{\text{eff}}^{x+}$; $\alpha$ is a prescribed parameter for regulating the whole dispersion extents of $\rho_{\text{eff}}^{x}$ and $K_{\text{eff}}$; $\rho_i$ represents the material phase in optimization and declares the air (0) or solid (1) attribute of a pixel; $CD_{\text{air}}$ denotes the number of the connected air domains in the microstructure; $\Sigma$ means the design domain shown in Fig. 1(c). Here we employ the simple geometrical constraint in Eq. (21) based on the fact that multiple connected air domains usually reduce the wave transmission and will form several closed cavities, resulting in extremely narrow-bandwidths of the resonances. Note that two special structural constraints in Eqs. (22) and (23) are introduced to make the optimized microstructure more reasonable in physics and manufacturing, respectively. Given that the AMMs with very narrow air channels usually incur the significant viscous losses which are induced by the near-wall viscosity effect (Cheng et al., 2015), it is necessary to restrict the dimensions of all air channels for circumventing this problem. That is, the minimum of the array $w_a$ composed by every air channel size should be larger than a pre-setting parameter $w_a^*$. Furthermore, the similar control over the solid components is also needed for topology optimization, especially for satisfying the sufficient strength requirement and fabrication of the metamaterial samples (Cheng et al., 2015; Dong et al., 2017, 2018). Therefore, we set the constraint in Eq. (23) to overcome these two issues, i.e., the minimum of an array $w_s$ including every solid size must be larger than an empirical value $w_s^*$.

Moreover, without taking any control measure, inducing the overlapping resonances can easily bring about two different variation performances for $\rho_{\text{eff}}^{x}$ and $K_{\text{eff}}$, making the broadband double negativity impossible. Therefore, we use the special physical constraint in Eq. (24) to precisely control the performances of $\rho_{\text{eff}}^{x}$ and $K_{\text{eff}}$ at the sampling frequency points. In terms of the discrete positive $\rho_{\text{eff}}^{x+}$ and $K_{\text{eff}}^{+}$, their possible variation with the increase of frequency can be generalized into two major categories: one is the simultaneous increasing tendencies (i.e., case 1); the other is the simultaneous non-increasing tendencies (i.e., case 2). Throughout this paper, we will demonstrate the crucial role of these two mechanism constraints for inducing two novel double-negative microstructures.

Obviously, the optimization problem in Eqs. (18)-(24) involves different kinds of constraints, intensifying the difficulties of optimization search in the large design space. Many topology optimization methods can effectively solve the various structure optimization problems in different fields (Zhou et al., 2011; Guo et al., 2014; Li et al., 2016; Aage et al., 2017; Zhang et al., 2018; Wang et al., 2018). Here, owing to the strong universality, the improved two-stage single-objective genetic algorithm (GA) (Dong et al., 2014a, 2014b, 2017, 2018) is utilized to solve the proposed optimization problem. GA treats the microstructure in $N_1 \times N_1$ pixels as a binary chromosome and mimics the evolutionary process by applying the natural selection principle to every generation towards the best design solution. First, an initial population of $N_p$ individuals is randomly generated. To improve the effectiveness of any microstructures, a special "abuttal entropy filtering" (Dong et al., 2014b) is applied for every microstructure to slightly fill up some isolated voids and remove some isolated elements. Secondly, every individual is evaluated for the fitness function and constrains. Then, GA uses the repetitive operators including the tournament selection, uniformed-matrix crossover and uniformed-matrix mutation to produce the offspring generation. Finally, the representative elitism strategy (Dong et al., 2014a) is utilized to improve and accelerate the optimization. After the prescribed number of generations, GA produces the optimized individual at the first stage. Introducing the optimized individual as the "seed" structure with $N_2 \times N_2$ pixels, GA are performed through the corresponding genetic operators at the second stage. Repeat all procedures generation by generation, and then bring about the final optimized microstructure towards to optimization problem in Eqs. (18)-(24).

## 3. Results and discussions

In this section, the proposed topology optimization formulation in Eqs. (18)-(24) is applied to design the square-latticed subwavelength metamaterial to obtain the broadband double negativity within a target frequency range [$\Omega_{min}$, $\Omega_{max}$]. Three representative symmetries including the square, chiral and orthogonal cases are investigated to give the synthetical and thorough insight into the beneficial topological feature of microstructures. All optimizations in this paper start from the random initial population. We adopt the following mass densities and speeds of sound for the air and solid materials: $\rho_{air}$=1.204 kg/m$^3$, $c_{air}$=343 m/s, $\rho_{solid}$=1230 kg/m$^3$ and $c_{solid}$=2230 m/s (Shen et al., 2016). For the optimization performances, unless otherwise indicated, the target frequency range is defined as [100 Hz, 8000 Hz]. The normalized frequency $\Omega=\omega a/2\pi c$ is introduced for convenience, where $a$ denotes the lattice constant, and $c$ is the acoustic velocity of the air. The normalized target range is [$\Omega_{min}$, $\Omega_{max}$]=[0.002476, 0.198061]. The number of sampling frequency points $M$ is set as 11, which is suggested by the numerical tests considering the computing cost and effectiveness of the discrete description. The solid constrained parameter is selected as $w_s^*$ =$a$/30 for all optimizations. The algorithm parameters of GA are the population size $N_p$=30, the crossover probability $P_c$=0.9, the mutation probability $P_m$=0.02, and the championship selection size $N_c$=18. At the first stage, the optimization is performed by 2500 generations in the 30×30 pixels. For the fine description of topologies, the optimization with the other 2500 generations is executed in the 60×60 pixels. Meanwhile, the optimized microstructure in the first stage is introduced as an initial design at the second stage. All two-stage optimization processes are implemented within 34.5 hours on a Linux cluster with Intel Xeon X5650 Core @ 2.66 GHz. The numerical simulations of dispersion relations, effective parameters, eigenstates, transmission spectrums and evanescent wave transmission are carried out by ABAQUS 6.14-1. The simulations of negative refraction and acoustic subwavelength imaging are accomplished by COMSOL Multiphysics 4.4. An acoustic experiment is conducted to demonstrate the subwavelength imaging of topology-optimized space-coiling AMM to exhibit the correctness of explored double-negative mechanisms and then show the potential of our topology optimization framework.

### *3.1. Optimized double-negative AMMs under simultaneous increasing tendencies of the effective parameters*

This subsection presents the optimization results with the prescribed simultaneous increasing tendencies ($\beta>0$) of the effective parameters, $\rho_{\text{eff}}^x$ and $K_{\text{eff}}$ (i.e., case 1 in Eq. (24)) for the normalized subwavelength target range [0.002476, 0.198061]. Some representative topological features, evolution history, various physical characterizations, negative properties of optimized metamaterials are analyzed and discussed in details. The typical and novel LC resonances contributing for the acoustic broadband double negativity are revealed through the resonance-cavity-based AMMs for the first time. All mentioned frequencies in the following contents refer to the normalized ones.

### 3.1.1. Topology-optimized resonance-cavity-based AMMs

#### 3.1.1.1 Square, chiral and orthogonal symmetries

We firstly design the microstructure with square symmetry and explore the effects of $\alpha$ and air channel width on optimized topologies and double negativity, as shown in Fig. 2(a). Extracting their macroscopic geometry features, it is interesting to find some common characteristics: (1) big air cavities connected through narrow air channels, and (2) big solid regions separated by the air domains. Like the Helmholtz resonator (Fang et al., 2006), multiple cavities in Fig. 2(a) can induce the negative effective bulk modulus. The distributions of hard solids can cause the large reflection with the limited space, which is beneficial for the occurrence of large refractive index. The above characteristics of resonance and large refractive index are dovetailed with the settings of the objective function in Eq. (19). With a larger $\alpha$, the metamaterial S2 has more cavities than S1. For the air-solid metamaterials with the viscous losses (Cheng et al., 2015), the widths of air channels have the appreciable impact on the efficacy of metamaterials. Fortunately, increasing this feature size can admittedly reduce the viscosity factor of metamaterials. To show the effect of air channels, we illustrate the optimized metamaterial S3 for the typical feature sizes of $w_a^* = a/15$ in Fig. 2(a). Compared with S1, S3 has the similar topological features except four additional slender hard solid plates. For clearly showing the desired negative properties, we present in Fig. 2(d) the double-negative ranges and quasi-static refractive index and impedance of the AMMs in Fig. 2(a). Since the effect of $\alpha$ is non-monotonic, we can only suggest its suitable range of [1.0, 1.5] in which double negativity can be effectively realized. The relatively larger refractive index usually makes the relatively wider double negativity possible. Combing the microstructure topologies, the variation of impedance shows that the reduction of air cavity domains can cause the increase of impedance. Hence S1 can keep the relatively large refractive index while maintain the reasonable wave transmission. The difference between S1 and S2 shows that the superabundant cavities may result in the narrow-band double negativity. The difference between S1 and S3 means that the wider air channels will evoke the smaller regions of air cavities, ultimately leading to the narrow-band double negativity.

To check the effectiveness of the proposed optimization formulation, we further investigate the topology optimization with the chiral and orthotropic symmetries, see Figs. 2(b) and 2(c). Figure 2(d) displays their double-negative ranges and quasi-static refractive indices and impedances. All chiral metamaterials in Fig. 2(b) contain a large air cavity in the center and four rotationally distributive solid blocks. For the unitcell domains marked by dashed lines, four corner regions of the metamaterial can be regarded as four small air cavities. With the same $w_a^*$, S5 has the better performance of double negativity than S4 and S6. So $\alpha$=1.0 is effective to balance the large refractive index and appropriate dispersion extent. Unlike the square-symmetry case in Fig. 2(a), the topology of S7 demonstrates that large $w_a^*$ prescribed in optimization can naturally thinning the solid components.

Moreover, the chiral symmetry is superior to the square symmetry if the complexity of structure is ignored. The double-negative ranges of optimized chiral metamaterials are apparently larger than those of the square-symmetry ones in Fig. 2(a). Therefore, multiple air cavities combined with the zigzag air channels and solid parts provide the ideal geometrical platform for the broadband double negativity.

Figure 2(c) presents the optimized orthogonal-symmetry metamaterials. For the subwavelength imaging, it is essential to make sure that the effective bulk modulus is isotropic during optimization for all potential designs. Although the effective bulk modulus is normally isotropic if the operating wavelength is larger than $5a$ (recall that $a$ is the lattice constant) (Popa et al., 2009; Xie et al., 2013), many highly complex orthogonal-symmetry metamaterials in optimization may have distinct behaviors along two principle directions, resulting in the high anisotropy and even possible coupling effects of effective bulk modulus using the present simulation model in Fig. 1(a). As a remedy approach, we force the relative difference between $K_{\text{eff}}$ retrieved from the $x$ and $y$ directions to be smaller than 5%. We can observe from S8 and S9 that the multiple air cavities and several isolated solid blocks become the main topology features for double negativity. Similar with the results in Fig. 2(a), increasing $\alpha$ tends to increase the numbers of the air cavities and solid blocks. The difference between S8 and S10 shows that the geometry would be simpler and has the thinner hard solid plates when the wider air channels are needed.

Through comparing the results in Fig. 2(d) for three cases, we can make the following observations. For the low-frequency property, the optimized AMMs have the good double negativity, with chiral symmetry the best and orthogonal symmetry the second. The behaviors of refractive index and impedance are positively correlated. The orthogonal-symmetry AMMs can realize the similar refractive index with the chiral-symmetry ones. But the chiral-symmetry AMMs have to face the relatively large impedance. Comparing the double-negative ranges of S4-S7, we find that the effect of $\alpha$ is smaller than the minimal air channel width. The similar feature can be observed from the orthogonal-symmetry case. Therefore, the control over the feature sizes of air channels should be a pivotal factor in designing the double-negative AMMs for practical applications. In the case of the same topological features, increasing the air channel widths will result in the decrease of refractive index and impedance for the optimization under simultaneous increasing tendencies. For the chiral symmetry, the relatively large (1.5) or small (0.5) $\alpha$ will result in the smaller refractive index. However, for the orthogonal symmetry, the relatively larger $\alpha$ can obtain the larger refractive index for double negativity.

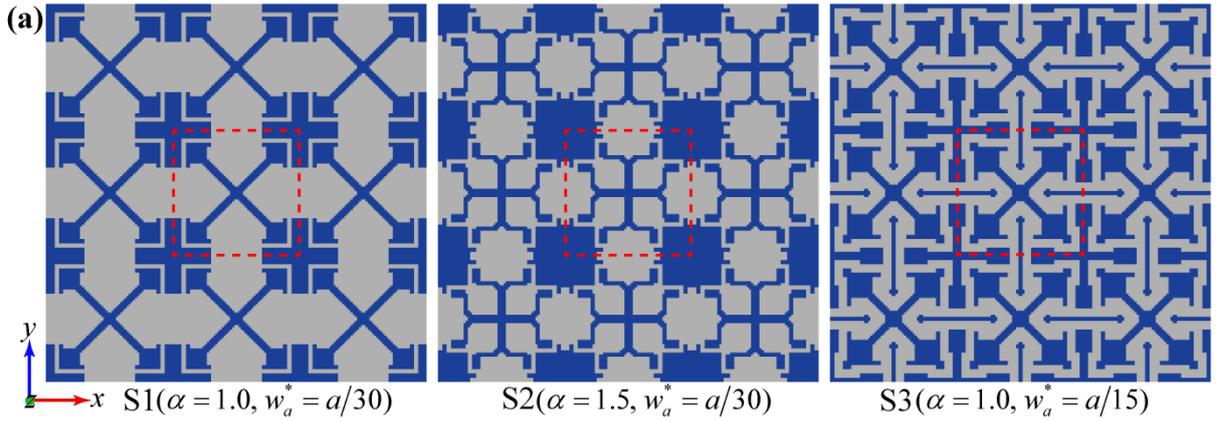

(a) S1($\alpha = 1.0$, $w_a^* = a/30$)    S2($\alpha = 1.5$, $w_a^* = a/30$)    S3($\alpha = 1.0$, $w_a^* = a/15$)

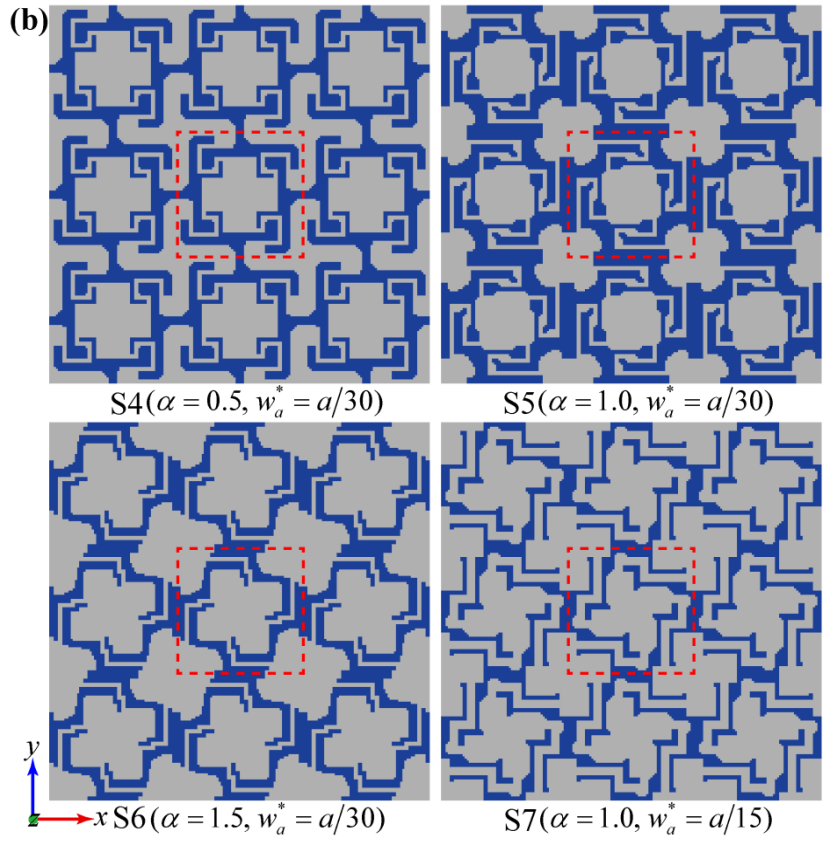
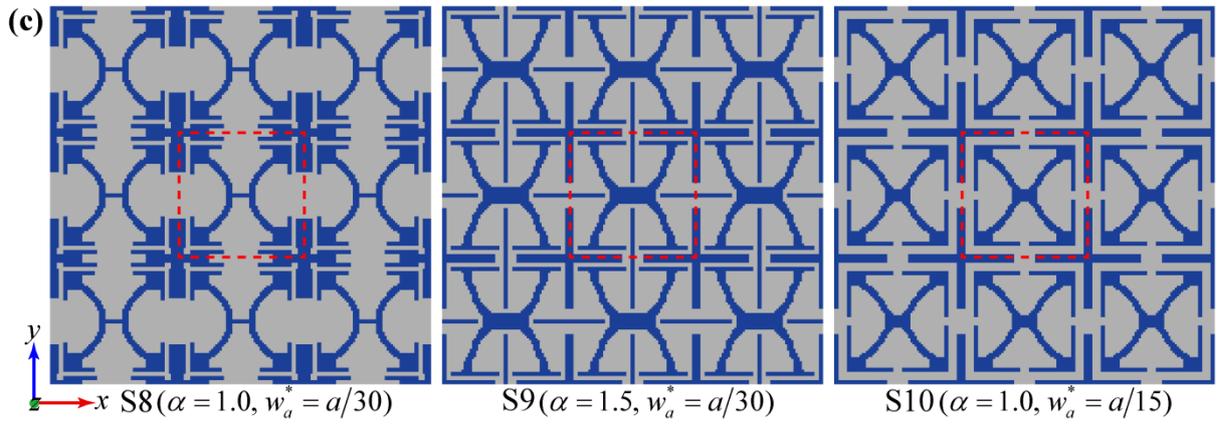

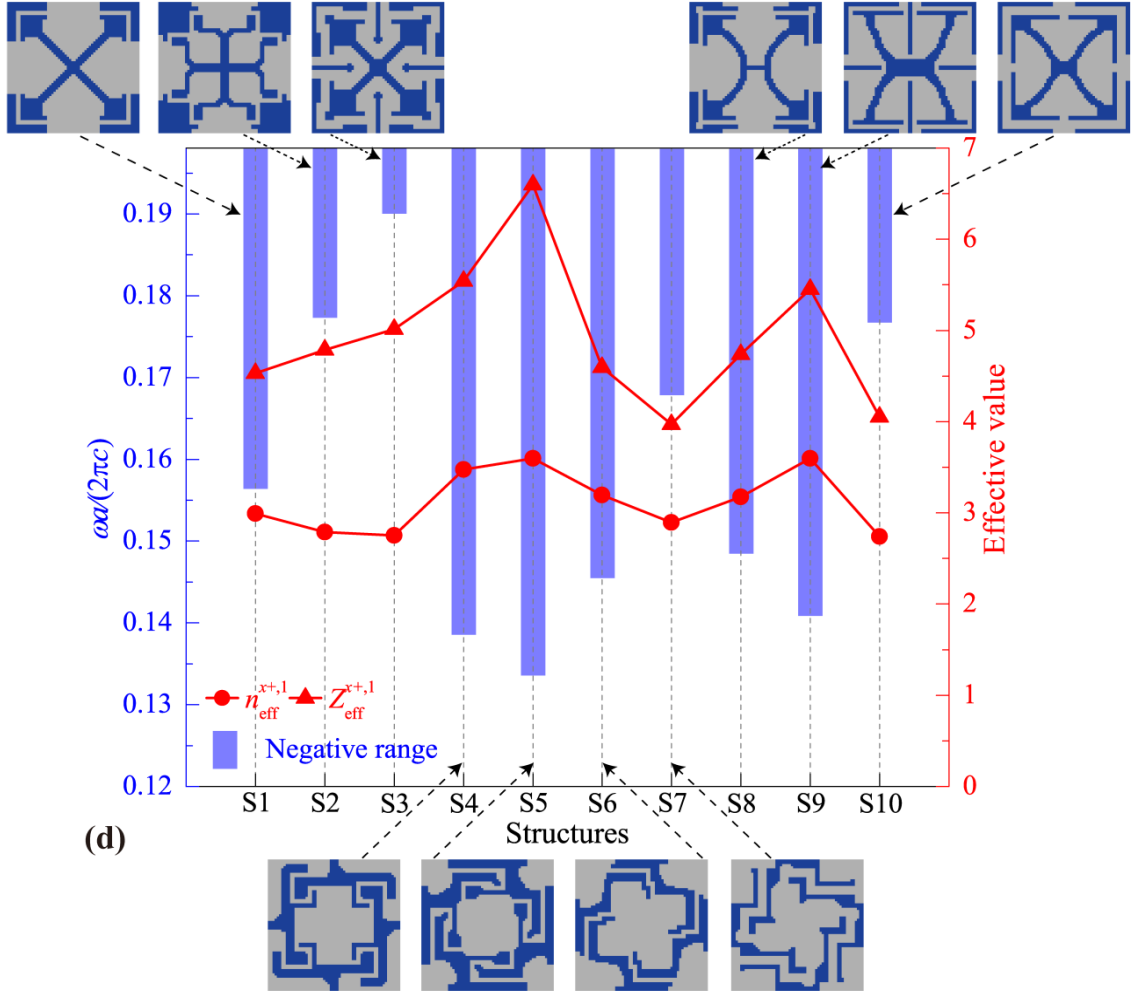

**Fig. 2. Topology-optimized resonance-cavity-based AMMs with three representative symmetries. (a)-(c)** Optimized metamaterials. All topology-optimized metamaterials are under the constraint of $\beta>0$. The target spectrum of S1-S10 is selected as [0.002476, 0.198061]. For the microstructures S8-S10, to guarantee the isotropy of effective bulk modulus, the relative difference between $K_{eff}$ retrieved from the $x$ direction and $y$ direction wave simulations are forced to be smaller than 5%. **(d)** Comparisons of the double-negative range, effective refractive index $n_{eff}^{x+,1}$ and impedance $Z_{eff}^{x+,1}$ for S1-S10. More performances are summarized in Appendix A.

Based on the solid-air system, the present topology optimization can effectively realizes the novel multi-cavities microstructures having ideal double negativity, and overcomes the limitations of single negativity of the Helmholtz metamaterials (Fang et al., 2006; Lee et al., 2010). From the prospective of double negativity, introducing the chirality is the best design approach; and the orthogonal symmetry takes the second place, followed by the square symmetry. Similarly, the chiral symmetry can induce the largest refractive index. From the prospective of topological features, three symmetric AMMs have the common ground for double negativity: the multiple air cavities, several solid blocks and relatively narrow air channels.

### 3.1.1.2 Analysis of representative AMM S1

In view of the most concise topological features and satisfactory double negativity, the AMM S1 is suitable to be systematically analyzed as the representative metamaterial. To clarify the evolution for the optimized topology of S1 in Fig. 2, Figure 3 shows the evolutionary history of the maximal fitness with the generation number during the "coarse to fine" optimization. Topology optimization starts from a randomly generated microstructure (G=0), which cannot satisfy the particular constraints of Eqs. (21)-(24). From the generation G=20 (F=0.1066) to G=215 (F=0.2563), GA can quickly capture the beneficial topological feature at the early evolution, i.e., four air cavities and two solid blocks. The maximal fitness change between G=215 (F=0.2563) and G=320 (F=2.6097) implies that the small central solids can contribute to the formation of double negativity. From the generation G=320 (F=2.6097) to G=3435 (F=3.7124), the microstructure turns to possess the longer air channels at four corners between the air cavities. Furthermore, the microstructure obtains the clearer edge descriptions and more smooth geometrical layouts. From G=3435 (F=3.7124) to G=4276 (F=3.7213), the slightly increased fitness demonstrates that the larger cavities should be the better choice under the circumstance with unchanged air channels. Therefore, we can generalize the beneficial topology features for the square-symmetry AMMs: four large enough air cavities, two independent solid blocks, the narrow enough air channels and several slender hard solid plates.

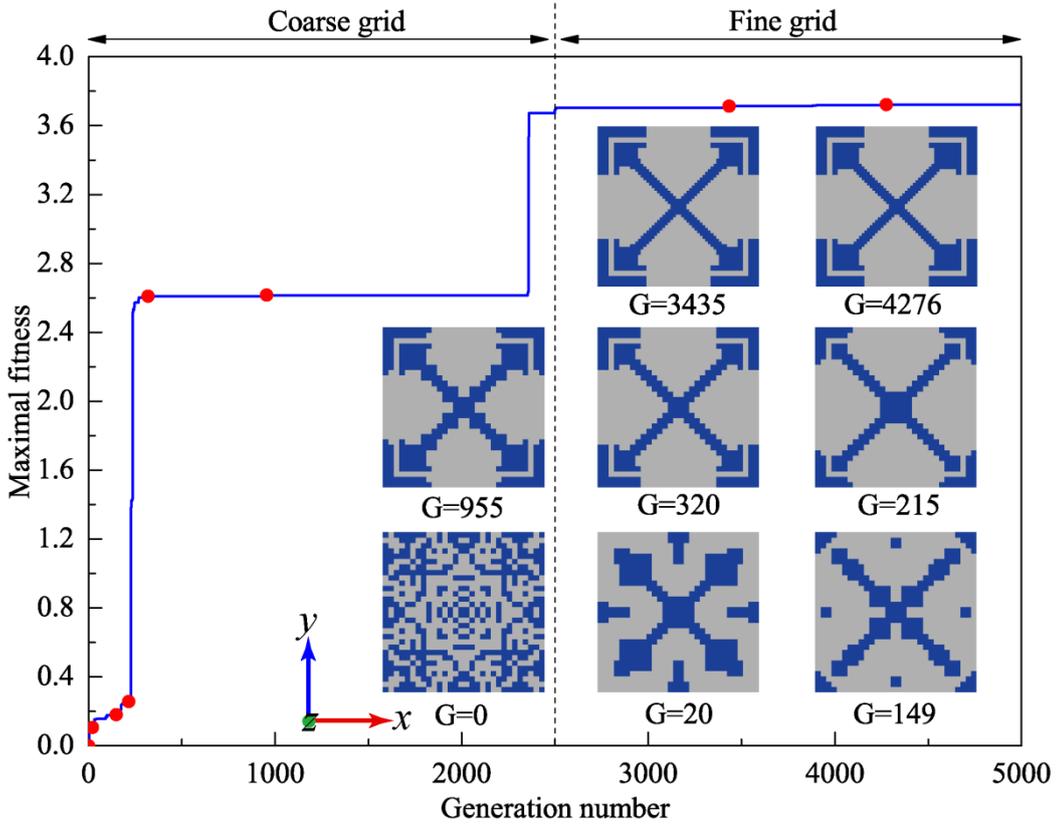

**Fig. 3. Evolutionary history for the generation of topology-optimized AMM S1 in Fig. 2.** Illustrations display eight representative topologies during the "coarse to fine" optimization. The objective function values of eight microstructures are 0 (G=0), 0.1066 (G=20), 0.1789 (G=149), 0.2563 (G=215), 2.6097 (G=320), 2.6149 (G=955), 3.7124 (G=3435) and 3.7213 (G=4276), respectively. The maximal fitness greater than or equal to 1.0 means emerging of double negativity.

To systematically characterize S1, we show in Fig. 4 the corresponding dispersion relations, effective constitutive parameters, wave transmission property and pressure magnification. Band structure in Fig. 4(a) displays a single band of [0.156384, 0.226167] with negative curvature. The near-linear trait of the first band indicates the homogenous wave behaviors in the deep-subwavelength scale. The effective parameters in Fig. 4(b) exactly show the simultaneous negative properties for $\rho_{\text{eff}}^x$ and $K_{\text{eff}}$ within the negative band range illustrated in Fig. 4(a). The positive values of $\rho_{\text{eff}}^x$ and $K_{\text{eff}}$ increase simultaneously at the established sampling frequency points (i.e., hollow circles and triangles). Indeed, their variations are consistent with the constraints required by Eq. (24). More importantly, $\rho_{\text{eff}}^{x}$ and $K_{\text{eff}}$ approach their resonance responses near the same frequency of 0.156384, and turn into the negative values simultaneously. This characteristic declares the best opportunity supplied by topology optimization for exploring the broadband double negativity. As plotted in Fig. 4(c), $n_{\text{eff}}^{x}$ and $Z_{\text{eff}}^{x}$ have the relatively large values and keep the simultaneous increasing variation as well. Negative $n_{\text{eff}}^{x}$ is generated within the same range as double-negative band in Fig. 4(a). There is a complete bandgap above the negative band, confirming the zero values of Re($n_{\text{eff}}^{x}$). Using the effective parameters, we can obtain in Fig. 4(a) the retrieved dispersion relations which perfectly match the band structures. The transmission spectrum in Fig. 4(d) shows that regardless of the thickness of AMM, the total transmission always appears near the lower edge of the double-negative range. In view of the dramatic change for Re($Z_{\text{eff}}^{x}$) near the lower-edge frequency, the perfectly-matched effective impedance Re($Z_{\text{eff}}^{x}$)=1 should be the physical origin of the total transmission. When the number of unitcell increases ($N$=1, 2, 10), the Fabry-Perot resonance conditions can be satisfied at more frequencies, consequently causing the more standing waves with high transmission compressed within the AMM.

Subsequently, wave transmission based on a microstructure S1 is calculated to demonstrate the essential resonances, see Fig. 4(e). Clearly, the localized pressure obtains two peaks with the increase of frequency, confirming the pressure magnification in the region of the upper cavity. To examine the effect of loss (Cheng et al., 2015), the simulations with the loss factor of 0.004, 0.0093 and 0.022 are performed and depicted in Fig. 4(e). Obviously, the viscous loss can only affect the magnification without incurring a shift in the operating frequency. This behavior demonstrates that S1 could keep the suitable balance between the resonance transmission and immunity to dissipation losses. Note that the specific resonance mechanisms will be analyzed and discussed in the following section.

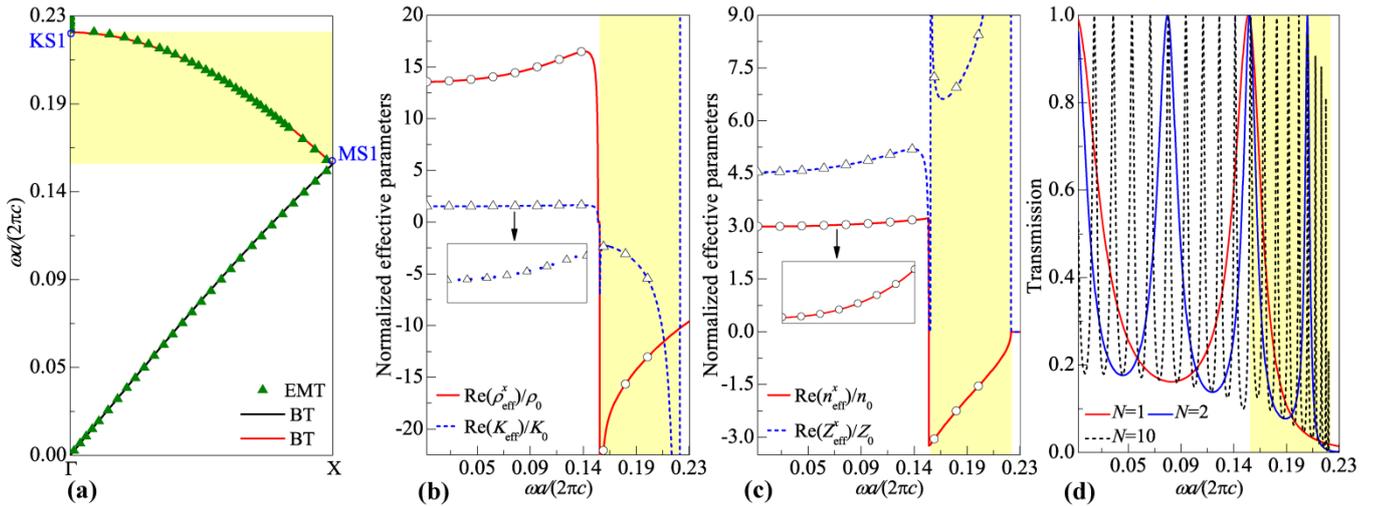

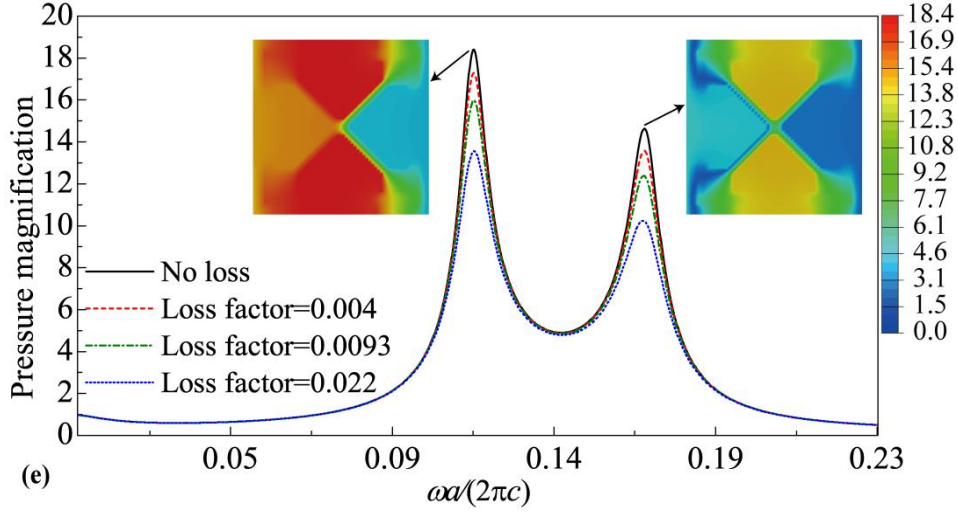

**Fig. 4. Various characterizations of topology-optimized AMM S1 in Fig. 2.** (a) Band structures along the ΓX direction (solid lines) based on the band theory (BT) and the retrieved dispersion relation Re(*k*)−*ω* (triangular scatters) based on the effective medium theory (EMT). (b) Relative effective mass density in the *x* direction and the bulk modulus. (c) Relative effective index and impedance along the *x* direction. All effective parameters in (b) and (c) are normalized to the background medium. (d) Transmission spectra of a finite AMM sample with different periodicities *N* along the *x* direction for the acoustic plane wave excitation. (e) Frequency dependence of pressure magnification in the region of upper cavity.

### 3.1.2. Mechanisms of optimized double negativity

To understand the physics of the double negativity in the resonance-cavity AMMs, we systematically study the eigenstates in the bands of S1-S6, S9 and S13. Figure 5 shows that optimized AMMs support typical LC resonances to guarantee the overlapping of different multipolar (monopolar, dipolar and quadrupolar, etc.) resonances. Eigenstates MS1 and KS1 give the understanding about the physical origin of negative band shown in Fig. 4(a). To be specific, MS2 shows the clear dipolar resonance caused by the highly localized energy in the left and right cavities. The infinite value of effective mass density in Fig. 4(b) indicates that MS2 is responsible for the negative effective mass density. As for KS2, most energy is localized in four cavities, forming the quadrupolar resonance and causing the infinite effective bulk modulus. And the negative value for $K_{eff}$ in Fig. 4(b) confirms the negative bulk modulus produced by KS2. In this case, the combination of dipolar and quadrupolar resonances can generate the double negativity. Since the range of negative effective bulk modulus is smaller than that of the negative effective mass density, the range of the negative effective bulk modulus dominates the bandwidth of double negativity. Similarly, eigenstates MS2 and KS2 also clearly exhibit the dipolar and quadrupolar resonances with different locations and occupied spaces of localized energy, respectively. Because the double negativity originates from the overlapping resonances, the double-negative range is determined by the common spaces which can support two kinds of resonances. Consequently, S2 has the smaller double-negative range than S1. Base on the same principle, the resonance space of eigenstate KS3 implies that S3 should have the smallest double-negative range for the square-symmetry case. Unlike the square-symmetry cases, however, MS5 and KS5 show the quadrupolar and hybridization of quadrupolar and monopolar resonances for the negative effective mass density and bulk modulus, respectively. Apparently, their overlapping makes sure the larger double-negative range. Eigenstate KS6 also shows the similar hybridization effect only with different chiral resonance spaces. Nevertheless, MS9 and KS9 indicate that in topology optimization, the orthotropic symmetry mainly alters the locations and topologies of four cavities, instead of the forms of resonances for double negativity.

To reveal the above LC-resonance mechanisms, we take S1 and S5 in Figs. 2 as examples and illustrate their equivalent physical models ES1 and ES5 in Fig. 5, respectively. In fact, the acoustic resonator can be analogous to an inductor-capacitor circuit for resonance properties (Fang et al., 2006), whose enclosed cavity acts as the capacitor; and relatively narrow air channels can undertake the inductor. Obviously, our optimized resonance-cavity-based AMMs can be equivalent to the combined system of several inductor-capacitor circuits. When the pressure variation occurs in the channels, the distributions of inductors and capacitors determine the excited forms of LC resonances. In other words, the model ES1 can support the eigenstates MS1 and KS1 under two kinds of excitations, yielding the negative mass density and bulk modulus, respectively. However, the model ES5 has the different topological feature, i.e., four capacitors locate in the corners and one capacitor in the center. On one hand, this distribution can induce the quadrupolar resonance for negative effective mass density. On the other hand, if five capacitors are excited simultaneously, the hybridization of quadrupolar and monopolar resonance can arise from model ES5 as well.

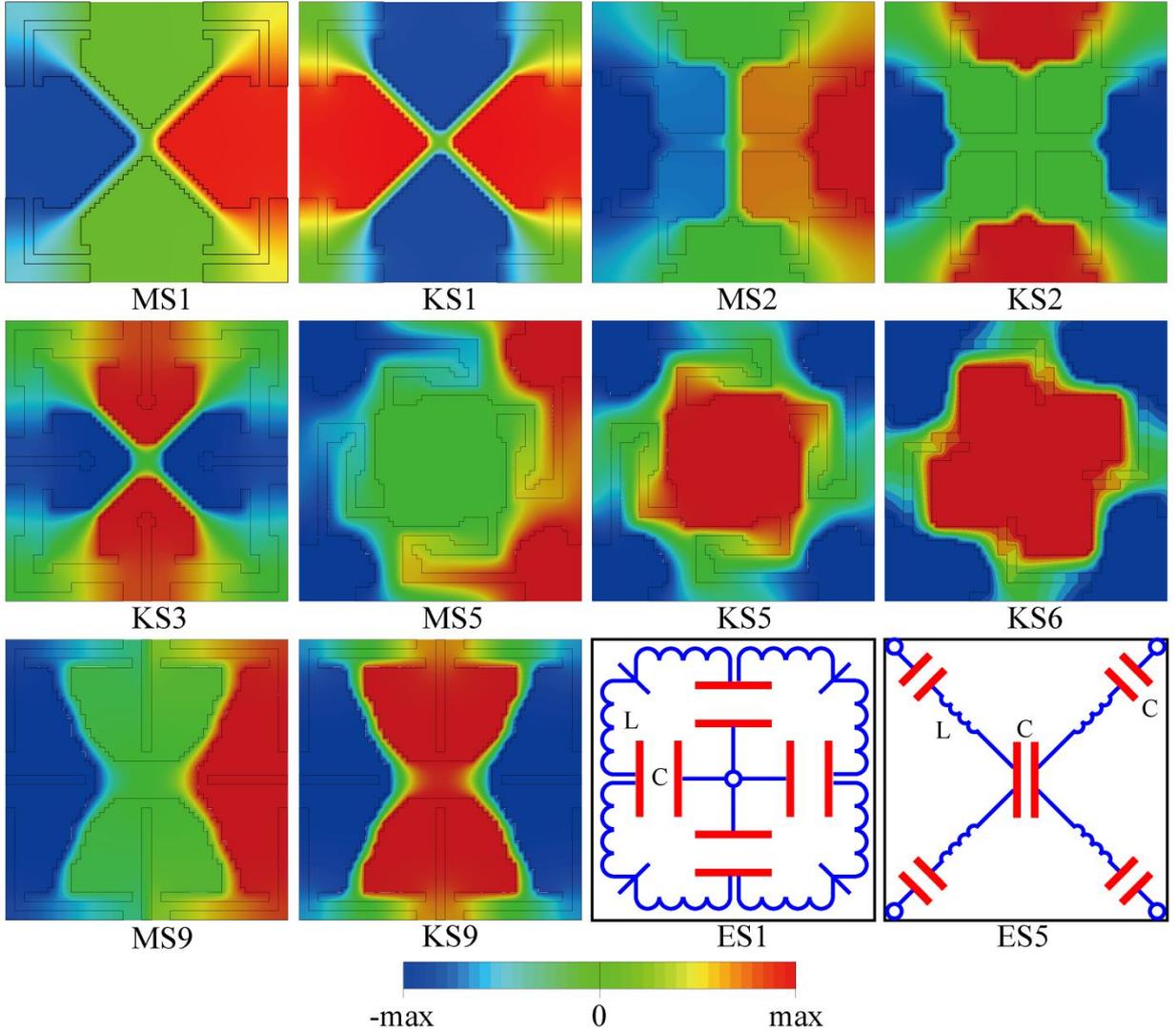

**Fig. 5. Specific eigenstates of topology-optimized AMMs S1–S3, S5, S6 and S9.** Eigenstates MS1 and KS1 are marked in Fig. 4(a). Eigenstates MS1 (S1), MS2 (S2), MS5 (S5) and MS9 (S9) correspond to the resonant modes, which induce the infinite effective mass density. Eigenstates KS1 (S1), KS2 (S2), KS3 (S3), KS5 (S5), KS6 (S) and KS9 (S9) correspond to the resonant modes, which induce the infinite effective bulk modulus. Sketches ES1 and ES5 represent the equivalent physical models of S1 and S5, respectively.

### 3.2. Optimized double-negative AMMs under simultaneous non-increasing tendencies of the effective parameters

This subsection presents the optimization results with the prescribed simultaneous non-increasing tendencies ($\beta \leq 0$) for both $\rho_{eff}$ and $K_{eff}$ (i.e., case 2 in Eq. (24)). To show the great potential of "non-increasing" mechanism, the optimization uses the more strict geometrical constraint, i.e., $w_a^*$ is set as a/15. Similarly, some representative topological features, evolution history, various physical characterizations, negative properties of the optimized metamaterials are analyzed and discussed in details. The emblematic and novel Mie resonances causing the acoustic broadband double negativity are perfectly revealed through the space-coiling AMMs.

### 3.2.1. Topology-optimized space-coiling AMMs

#### 3.2.1.1 Square, chiral and orthogonal symmetries

Since our proposed optimization framework in Eqs. (18)-(24) is general, we apply it to design the broadband double-negative AMMs under the simultaneous non-increasing tendencies ($\beta \leq 0$) considering the square, chiral and orthotropic symmetries, see Fig. 6. Interestingly, these AMMs exhibit the common topological features essentially different from those shown in Fig. 2: (1) several separated solid blocks in the coiling up space, (2) zigzag air channels forming the labyrinth layouts, and (3) a number of local air regions. Intuitively, like the extreme metamaterials reported by Liang et al. (2012), the waves can freely propagate in the curled space with materials of negligible loss, arousing the large phase delays within a small space and then realizing the large refractive index. Consequently, the band folding (Liang et al., 2012) supporting the double negativity should emerge at the subwavelength scale. Compared with the AMMs in Fig. 2, S11 has more separated solid blocks and complicated air channels, achieving the large effective refractive index without the resonance cavities. Note that S11 can only generate the double negativity within [0.25438, 0.379423], see Appendix A. This means that, with the square-symmetry assumption, the better driving force for the low-frequency double negativity should be the simultaneous increasing tendencies ($\beta > 0$). However, with the same $w_a^*$, the chiral- and orthotropic-symmetry AMMs in Fig. 6 have the wider double negativities at the lower frequency ranges. Consequently, the simultaneous non-increasing tendencies ($\beta \leq 0$) are more beneficial for the low-frequency broadband double negativity than the increasing tendencies ($\beta > 0$). In addition, both the optimized chiral- and orthotropic-symmetry AMMs can accomplish the preferable double negativity. This testifies the robustness of space-coiling topology for generating double negativity on the other side. The comparison of S12 and S13 shows that increasing $\alpha$ can reduce the air paths and induce the thinner hard solid plates. Since S14 is generated with the different target spectrum, the similar topology and double negativity of S14 demonstrate that the proposed optimization strategy is robust for the specific frequencies. From the results in Fig. 6(b), we can find that the wide air regions except the interconnection core regions and relatively thick hard solid plates can result in the degenerations of double negativity. Therefore, the most beneficial space-coiling topology should include the suitable zigzag channels, thin curved hard solid plates and interconnection core regions in the centers.

To make sure the negative properties of the space-coiling AMMs in Fig. 6(a), Figure 6(b) presents their double-negative ranges and quasi-static refractive index and impedance. For the low-frequency performance, the optimized AMMs, S12, S14, S15 and S16, have the conspicuous broadband double negativity transcending the previous extreme metematerials (Liang et al., 2012). Moreover, all optimized space-coiling AMMs in Fig. 6 possess the larger air channels than those of the previous extreme metematerials (Liang et al., 2012; Xie et al., 2013), thus obviously reducing the viscous loss. In spite of the more strict constraints of air channels than those in Fig. 2, the optimized AMMs in Fig. 6 can also obtain the desired negative bands. With the same symmetry, the variations of impedance are positively related to refractive index. Differences between S12, S14, S15 and S16 show that the orthotropic-symmetry AMMs can generate the similar refractive index with the chiral-symmetry ones. However, the corresponding impedance will be markedly increased. This further illustrates that the chiral symmetry can realize the ideal double negativity and wave transmission simultaneously. For the chiral symmetry, the difference between the performances of S12 and S13 shows that the relatively large $\alpha$ can induce the smaller refractive index and impedance, causing the relatively small double-negative range. In the other words, the variation of refractive index dominates the optimization in this case. However, for the orthotropic symmetry, the relatively large $\alpha$ mainly reduces the dispersion extents of $\rho_{\text{eff}}^x$ and $K_{\text{eff}}$, resulting in a smaller double-negative range.

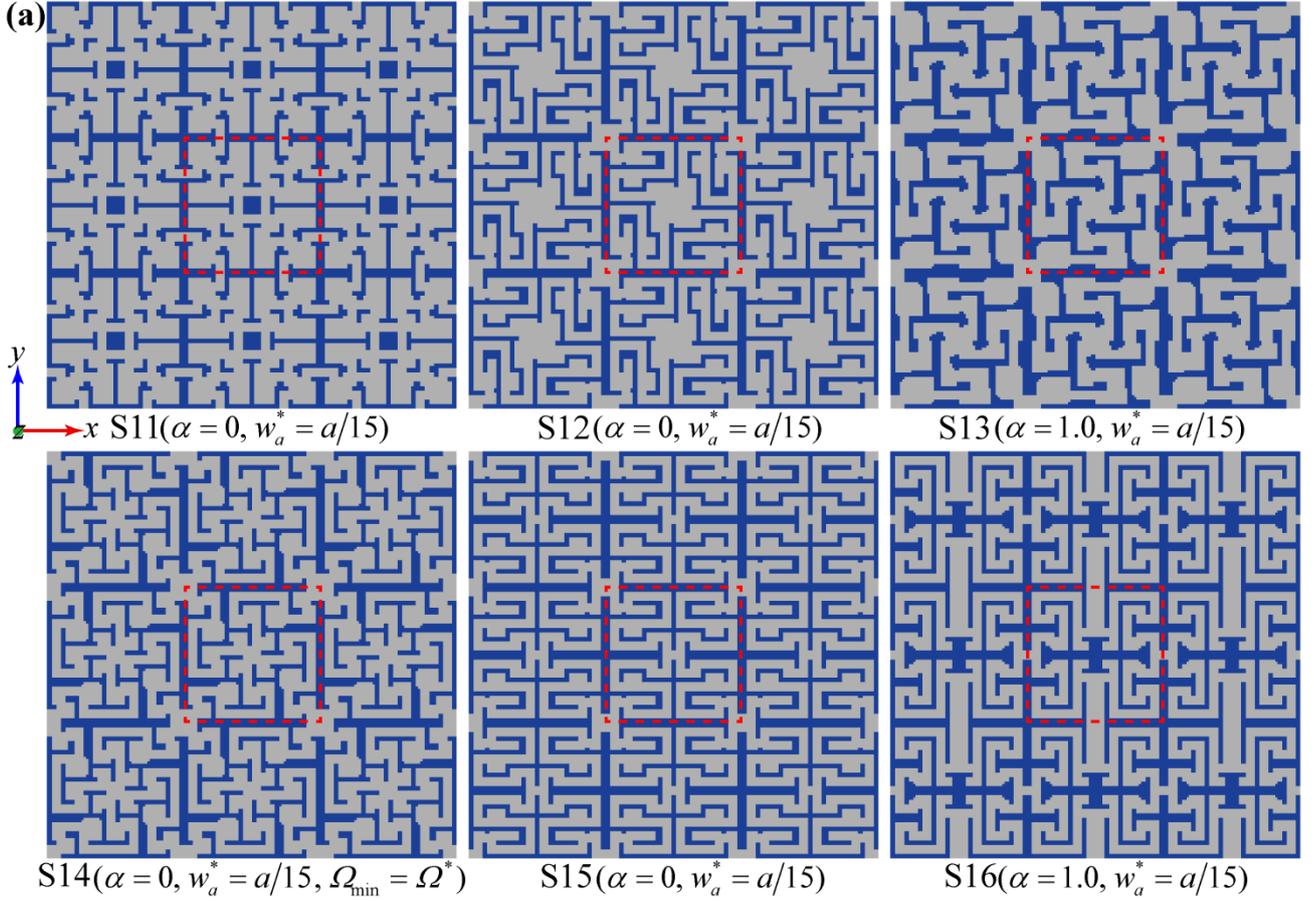

(a)

S11($\alpha = 0, w_a^* = a/15$)  S12($\alpha = 0, w_a^* = a/15$)  S13($\alpha = 1.0, w_a^* = a/15$)

S14($\alpha = 0, w_a^* = a/15, \Omega_{\min} = \Omega^*$)  S15($\alpha = 0, w_a^* = a/15$)  S16($\alpha = 1.0, w_a^* = a/15$)

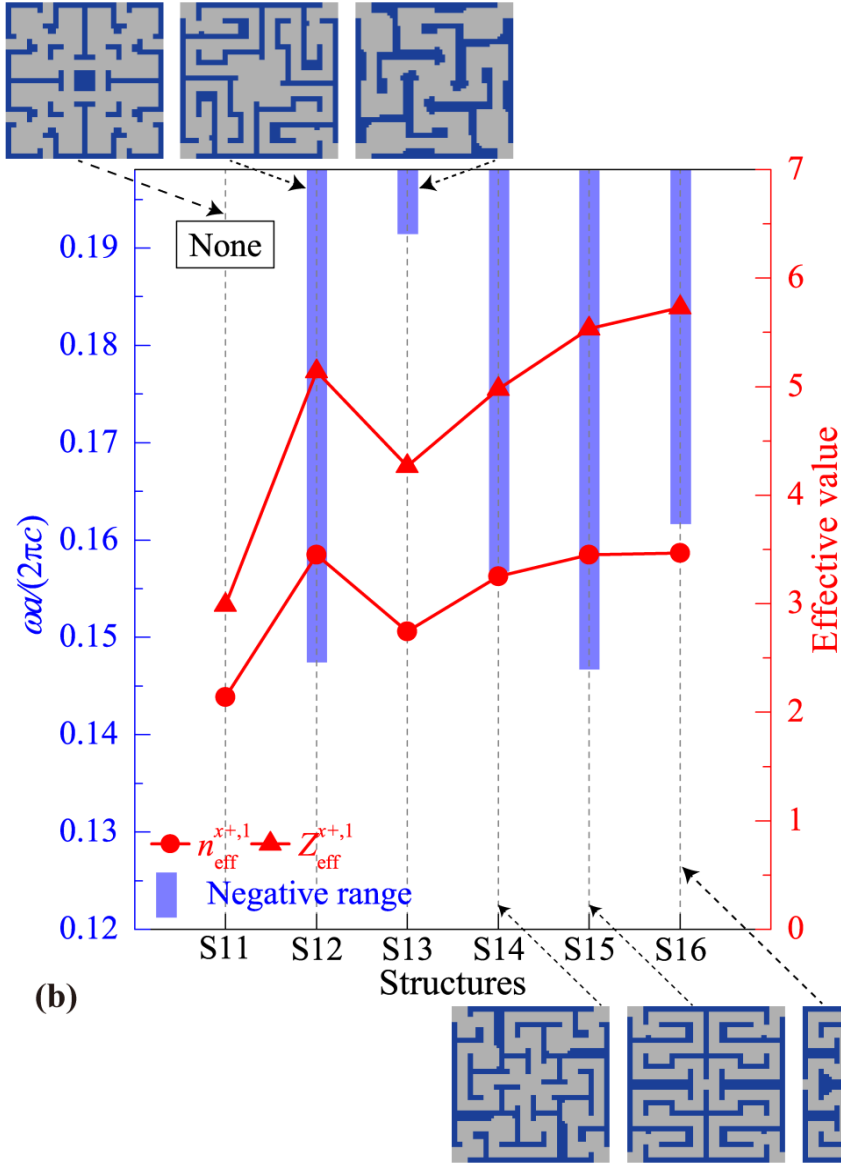

**Fig. 6. Topology-optimized space-coiling AMMs with three representative symmetries.** (a) Optimized microstructures. All topology-optimized metamaterials are under the constraint of $\beta\lesssim 0$. The target spectrum of S14 is selected as [0.09903, 0.198061] (i.e., $\Omega^*$=0.09903. The ranges of the other structures are set as [0.002476, 0.198061]. For the microstructures S15 and S16, to guarantee the isotropy of effective bulk modulus, the relative difference between $K_{\text{eff}}$ retrieved from the $x$ direction and $y$ direction wave simulations are forced to be smaller than 5%. (b) Comparisons of double negativities, the effective refractive index $n_{\text{eff}}^{x+,1}$ and impedance $Z_{\text{eff}}^{x+,1}$ for S11-S16. More performances are summarized in Appendix A.

With the prescribed typical symmetries, the present topology optimization can effectively realize the novel space-coiling microstructures having ideal double negativity, which breaks the restrictions of single negativity (Cheng et al., 2015) or narrow-band double negativity of the labyrinth metamaterials (Liang et al., 2012). From the prospective of double negativity, both the chiral and orthogonal symmetries provide the fantastic design approach, being far superior to the square symmetry. From the prospective of topological features, three symmetric AMMs have the common ground for negative properties: the suitable zigzag channels, thin curved hard solid plates and interconnection core regions in the centers. Regarding the manufacturing difficulty, it is easy to fabricate all space-coiling microstructure that are benefited from their lots of straight hard solid plates.

*3.2.1.2 Analysis of representative AMM S12*

To understand the origin of the space-coiling topology shown in Fig. 6, we show the evolutionary history of the maximal fitness with the generation number for the AMM S12 in Fig. 7. The noteworthy change from the generation G=0 (*F*=0) to G=10 (*F*=0.2859) illustrates the strong searching ability of GA which can find out the separated thick hard solid plates and interconnection core regions. The creation of double negativity at generation G=117 (*F*=1.5521) means that making channels and solids parts more curved is the effective approach for getting double negativity. From the generation G=117 (*F*=1.5521) to G=4680 (*F*=4.0818), the solid components become thinner while the curved extent of zigzag paths increases, thus producing the bigger double negativity range. Meanwhile, the interconnection core region in the center keeps a relatively large size. In spite of the similar space-coiling topology with the extreme metamaterials reported by Liang et al. (2012), the AMM S15 shows the emblematic features of larger interconnection regions and more curved degree of the hard solid plates, which gives rise to the better double-negative property.

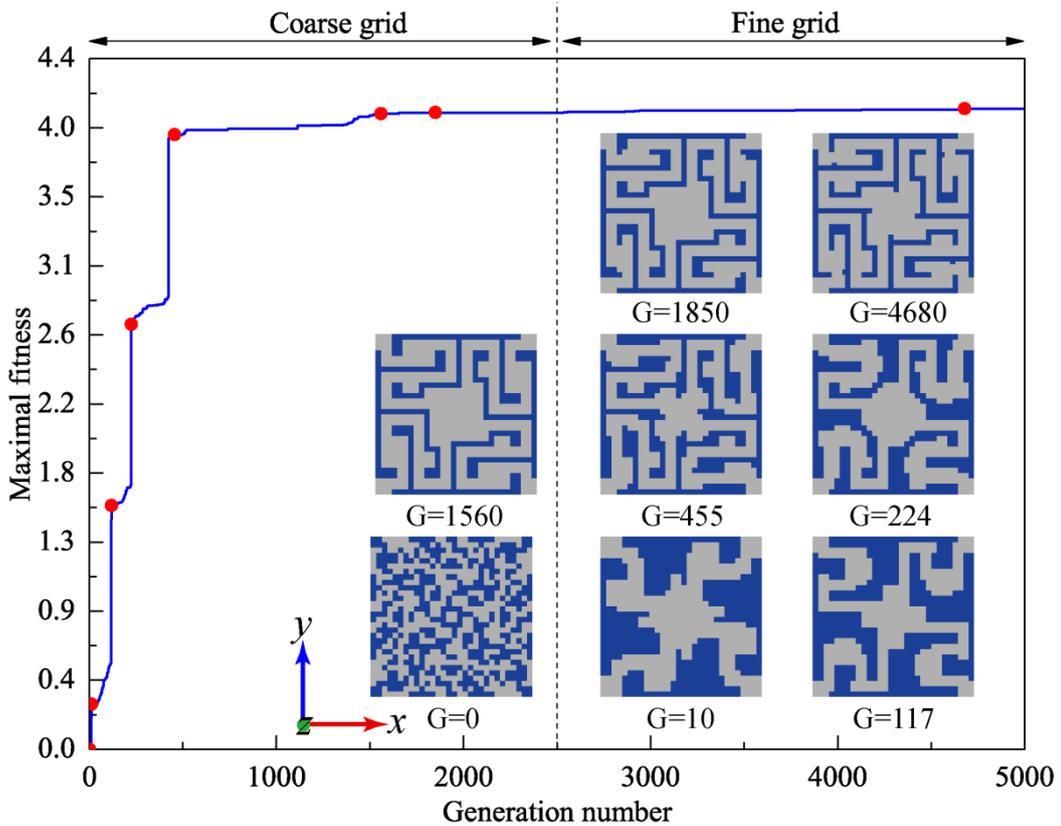

**Fig. 7. Evolutionary history for the generation of topology-optimized AMM S12 in Fig. 6.** Illustrations display eight representative topologies during the "coarse to fine" optimization. The objective function values of eight microstructures are 0 (G=0), 0.2859 (G=10), 1.5521 (G=117), 2.7063 (G=224), 3.9155 (G=455), 4.0487 (G=1560), 4.056 (G=1850) and 4.0818 (G=4680), respectively.

To characterize the double negativity of S12, we systematically study the dispersion relations, effective parameters and transmission spectrum, see Fig. 8. For the second band range displayed in Fig. 8(a), the different physical quantities coincide mutually. Interestingly, owing to the band folding, the slopes around the Γ point in both the ΓX and ΓM directions are almost the same at the first, second, forth, sixth, eighth and tenth bands. In fact, this band folding not only indicates the isotropic indices, but also gives rise to the ideal negative properties at both subwavelength and long-wavelength regimes. Just as the constraints of simultaneous non-increasing tendencies in Eq. (24), the positive $\rho_{\text{eff}}^{x}$ and $K_{\text{eff}}$ decrease simultaneously below the negative range. The simultaneous negative $\rho_{\text{eff}}^{x}$ and $K_{\text{eff}}$ in Fig. 8(b) and negative $n_{\text{eff}}^{x}$ clearly demonstrate the double-negative property of the negative band depicted in Fig. 8(a). Unlike the results in Fig. 8(b), the positive $n_{\text{eff}}^{x}$ and $Z_{\text{eff}}^{x}$ have the opposite variation patterns. Using the effective parameters in Figs. 8(b) and 8(c), the retrieved dispersion relations based on EMT can perfectly match the band structures. Benefitting from the perfectly-matched effective impedance Re($Z_{\text{eff}}^{x}$)=1, the total transmission appears near the lower edge of double-negative range whether the metamaterial layer is thin (N=1, 2) or thick (N=10), see Fig. 8(d). Besides, the high transmission at the other frequencies can also be obtained because of the satisfied Fabry-Perot resonance conditions.

The wave transmission based on a microstructure of S15 is further calculated to demonstrate the essential resonances, as shown in Fig. 8(e). Clearly, the pressure with large magnitude is mainly localized in the central interconnection region, showing the pressure magnification in the region of the upper cavity. Obviously, the viscous loss can only affect the magnification without introducing a shift in the resonance frequency. This behavior demonstrates that S15 could keep the suitable balance between resonance transmission and immunity to dissipation losses. Note that the specific resonance mechanisms will be analyzed and discussed in the following section.

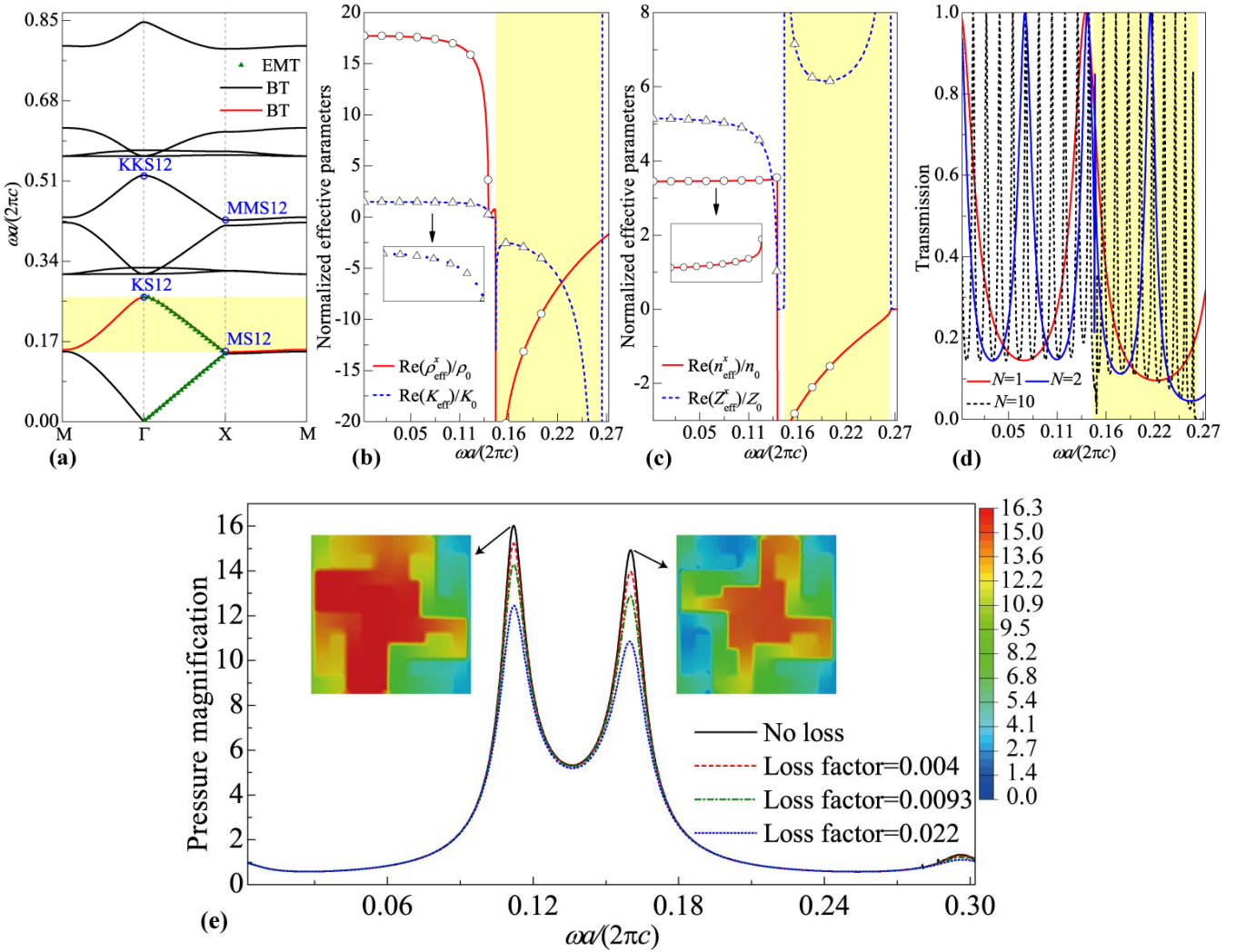

**Fig. 8.** Various characterizations of AMM S12 in Fig. 6. **(a)** Band structures (solid lines) based on the band theory (BT) and the retrieved dispersion relation Re($k$)−$\omega$ (triangular scatters) based on the EMT. **(b)** Relative effective mass density in the $x$ direction and the bulk modulus. **(c)** Relative effective index and impedance along the $x$ direction. All effective parameters in (b) and (c) are normalized to the background medium. **(d)** Transmission spectra of a finite AMM sample with different periodicities $N$ along the $x$ direction for the acoustic plane wave excitation. **(e)** Frequency dependence of pressure magnification in the region of central interconnection core.

### *3.2.2. Mechanisms of optimized double negativity*

To reveal the origin of the double negativity in the space-coiling AMMs, we take four metamaterials S12-S15 as examples and carefully investigate their specific eigenstates in Fig. 9. Obviously, eigenstates MS12 and KS12 portray the quadrupolar and hybridization of quadrupolar and monopolar Mie resonances which can essentially induce the negative effective mass density and bulk modulus, respectively. It is emphasized that these artificial Mie resonances distinctly differ from the LC resonances shown in Fig. 5. In principle, the Mie resonance usually appears in the structure having the high refractive index relative to the background medium (Brunet et al., 2015; Cheng et al., 2015). Instead of highly localizing all energy within several cavities, the Mie resonances can make the energy concentrate in the air regions while accompanying the apparent radiations, showing the feature of resonance scattering. Eigenstates MMS12 and KKS12 behave as the second-order quadrupolar and hybridization of the quadrupolar and monopolar Mie resonances, and characterize the double-negative essence of the sixth band in Fig. 8(a). Similarly, the high-frequency negative band (tenth band) in Fig. 8(a) also has the double negativity resulted from the higher-order Mie resonances. Eigenstates MS13 and KS13 also show that the double negativity of S13 is generated by the quadrupolar and hybridization of the quadrupolar and monopolar Mie resonances. Comparing the field distributions of MS12, KS12, MS13 and KS13, we can infer that the relatively larger interconnection core region will causes the double negativity at the lower frequency range, see the negative ranges of S12 and S13 shown in Fig. 6(b). In addition, eigenstates MS14 and KS14 shows that the Mie resonances can also be induced by the more complex labyrinth structure. This means that the space-coiling topology is very robust for generating the multipolar Mie resonances. In addition, eigenstates MS15 and KS15 show that the ortho-symmetric labyrinth topology is conducive to the similar Mie resonances as well. Hence the AMMs presented in Fig. 6 provide unanticipated topology features for both Mie resonances and double negativity. Note that the bandwidth of the double negativity is determined by the size of overlapping regions for two resonances. However, in case of the quadrupolar and hybridization of the quadrupolar and monopolar Mie resonances, the largest overlapping regions should be the four corner areas, which indicates the corresponding limited bandwidth. In particular, it is the combination of multipolar resonances that provides the broadband double negativity over the previous studies of space-coiling AMMs (Liang et al., 2012). Overall, eigenstates in Fig. 9 disclose that the optimal mechanism with space-coiling topology for broadband low-frequency double negativity should be the combination of quadrupolar and hybridization of quadrupolar and monopolar Mie resonances.

To reveal the above Mie-resonance mechanisms, we also present in Fig. 9 the equivalent physical models ES12 and ES13 of the AMMs S12 and S13, respectively. Because S12 has the high effective refractive index and an air cavity in the center, the whole microstructure can be equivalent to four channels composed of the ultraslow medium connected with an air interconnection core. And they are separated by the solid frame materials. When the waves propagate in the four channels with different forms of phases, the model ES12 can produce the quadrupolar Mie resonance MS12 or the hybridization of the quadrupolar and monopolar Mie resonance KS12. In particular, the relatively large interconnection core contributes to this hybridization. Since the holistic effective refractive index of S13 is smaller than that of S12, the corresponding equivalent model ES13 has the different straightened channels while ensuring the similar geometrical feature. In addition, since S13 has the smaller air cavities in the center than S12, ES13 should have the smaller air interconnection core as well. As a result, the difference between ES12 and ES13 mainly affects the frequency range of double negativity.

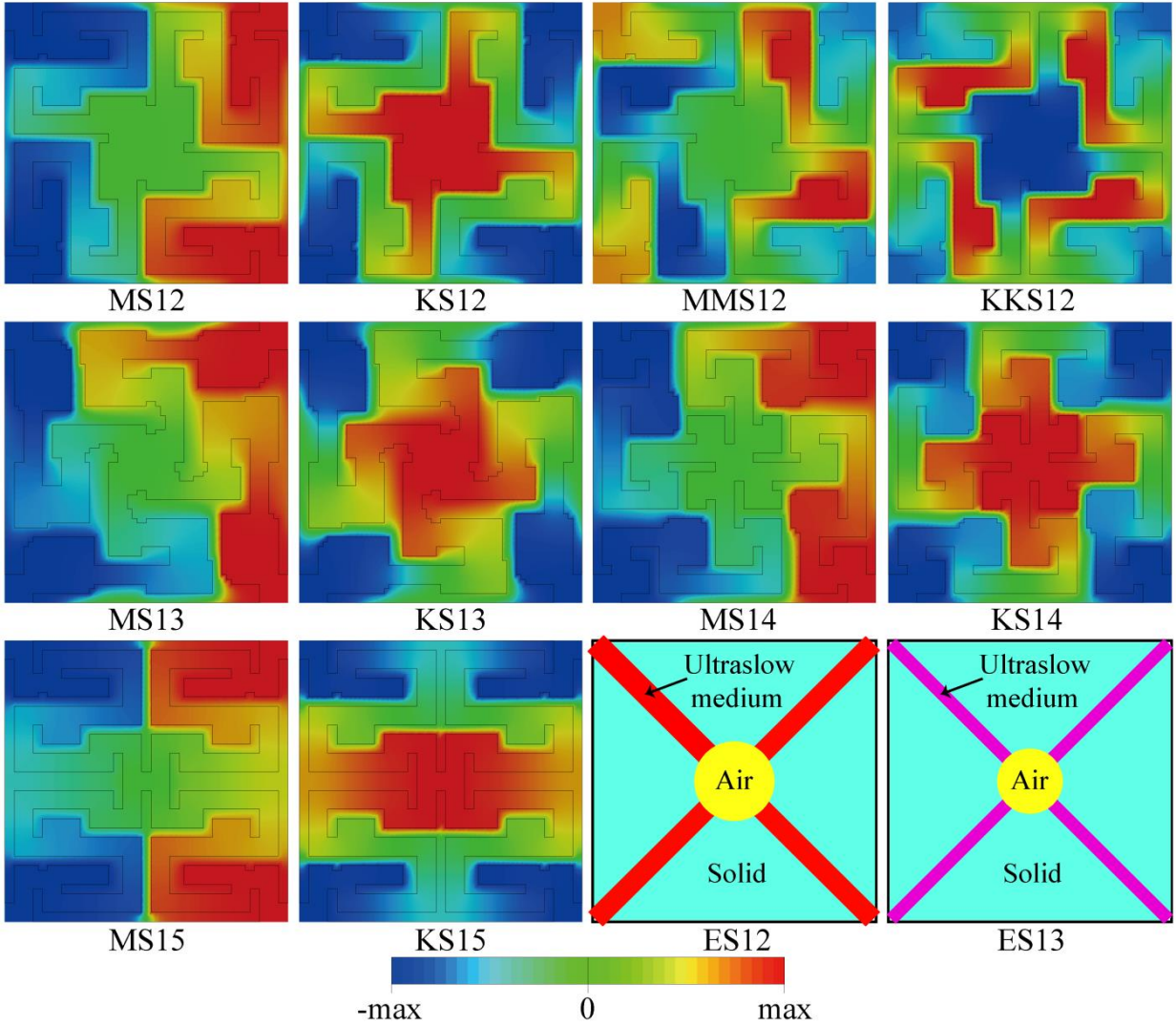

**Fig. 9. Specific eigenstates of topology-optimized AMMs S12-S15.** Eigenstates MS12 and KS12 are marked in Fig. 8(a). Eigenstates MS12 (S12), MMS12 (S12), MS13 (S13), MS14 (S14) and MS15 (S15) correspond to the resonant modes, which induce the infinite effective mass density. Eigenstates KS12 (S12), KKS12 (S12), KS13 (S13), KS14 (S14) and KS15 (S15) correspond to the resonant modes, which induce the infinite effective bulk modulus. Sketches ES12 and ES13 represent the equivalent physical models of S12 and S13, respectively.

### 3.3. Brief summary on two categories of AMMs

Due to the tremendous inverse-design ability, topology optimization has explored two categories of novel AMMs with the broadband double negativity, namely the resonance-cavity-based and space-coiling metamsaterials, respectively. For the desired negative properties, the space-coiling metamaterials can realize the wider double negativity within the lower frequency ranges, showing the bigger superiority than the resonance-cavity-based ones. In addition, the double negativity of the space-coiling metamaterials is less affected by the width of the air channels. And space-coiling metamaterials can produce the double negativity by introducing the larger air channels. The beneficial resonance-cavity-based metamaterials should possess the multiple air resonance cavities, several hard solid plates and air channels. For the space-coiling metamsaterials, the beneficial topologies should include the suitable zigzag channels, thin curved hard solid plates and interconnection core regions in the centers. For the double-negative mechanisms, the resonance-cavity-based metamaterials benefit from the novel multipolar LC resonances. But the space-coiling metamaterials supports the novel multipolar Mie resonances. In principle, both novel mechanisms overcome the limitations of the reported negative resonances (Fang et al., 2006; Lee et al., 2010; Liang et al., 2012; Xie et al., 2013; Cheng et al., 2015), representing the optimal physical essences for broadband double negativity so far. Because of most straight hard solid plates, the space-coiling metamaterials are easier to be manufactured than the resonance-cavity-based ones. Anyway, the topology-optimized AMMs presented in this paper can achieve the double negativity in a brand-new structural style.

### 3.4. Potential applications using topology-optimized double-negative AMMs

This subsection presents the numerical results of broadband double negativity and the enhancement of evanescent wave transmission for LC-resonance and Mie-resonance optimized AMMs. Then we respectively show the negative refraction and acoustic subwavelength imaging with the high transmission. Finally, the experimental demonstration of subwavelength imaging is successfully realized.

### 3.4.1. Numerical demonstrations of wave behaviors

To demonstrate the desired negative dispersions, we illustrate the equi-frequency surfaces (EFSs) of metamaterials S1 and S12 in Figs. 10(a) and 10(b), respectively. It is noted that two negative bands show the quite isotropic behavior within the whole range except the lower edge of the bands with the slight anisotropy. In addition, we can clearly observe the striking difference between the LC-resonance and Mie-resonance negative bands in Figs. 4(a) and 8(a), i.e., two bands are arc-shaped and near-straight, respectively. The variations of the EFSs with the frequency increasing in Figs. 10(a) and 10(b) mean that the negative group velocities should occur along all directions. When waves are incident to the interface between the AMMs and background media, the refracted group velocity should be pointed to the direction of frequency increasing which is perpendicular to the contours, causing the expected negative direction. It is worth mentioning that the target frequency spectrum of [0.002476, 0.198061] can guarantee the all-angel negative refraction for the whole target frequency spectrum. Then the subwavelength imaging can be realized within the whole target frequency spectrum as well.

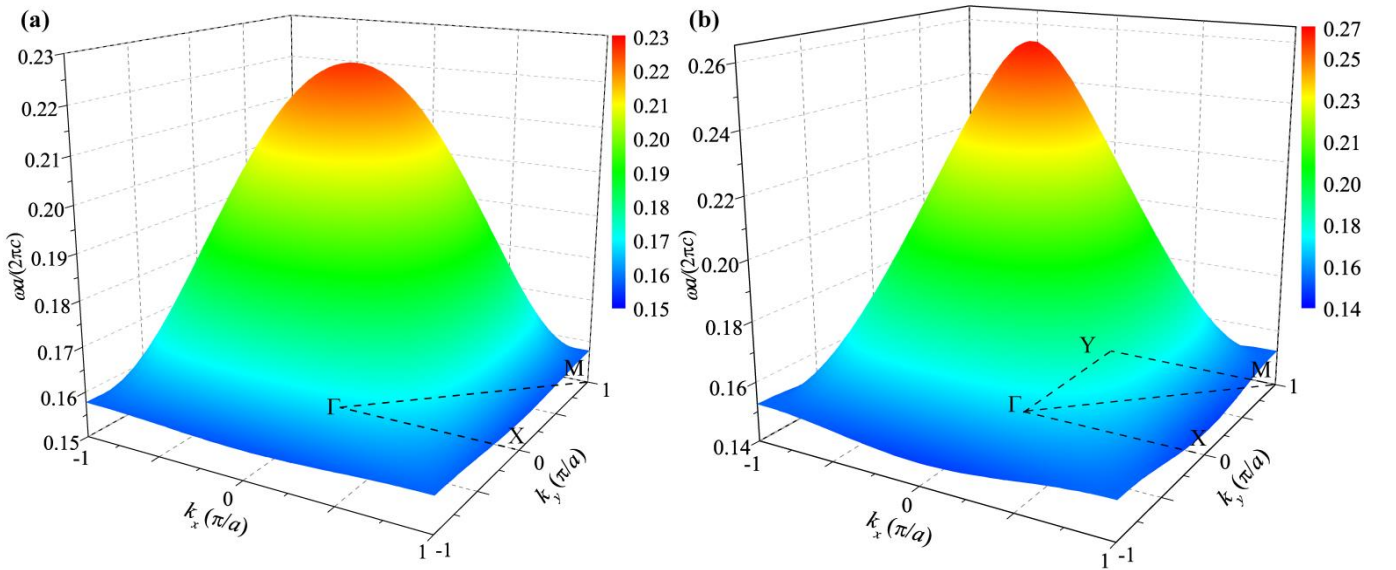

**Fig. 10. Equi-frequency surfaces of topology-optimized AMMs.** Surface plots of the first negative band (the second band) over the whole Brillouin zone for S1 **(a)** and S12 **(b)**.

To validate the negative refraction and subwavelength imaging, we display the corresponding simulation results for the metamaterials S1 and S12 in Figure 11. As predicted in Figs. 11(a) and 11(b), when the Gaussian beam (45°) of an acoustic wave is incident from the left region, the desired negative refraction with high transmission can be clearly observed at $\Omega$=0.173303 and $\Omega$=0.160925, respectively. Meanwhile, when a point source is excited in the left of the metamaterial slab, the obvious imaging effect for S1 and S12 occurs in the exiting surfaces of the slabs in Figs. 11(c) and 11(d). Their full widths at the half maximum (FWHM) of images are 0.44$\lambda$ and 0.39$\lambda$, respectively, which beyond the diffraction limit. Therefore, topology-optimized metamaterials S1 and S12 are demonstrated to have ability of realizing the subwavelength imaging. In addition, we also present the simulations for S12 at the different frequencies in Figs. 11(e)-11(g). The comparison between the results in Figs. 11(d)-11(g) shows that the metamaterial lens can effectively generate the stable subwavelength imaging within the negative range. With the increase of frequency, the imaging resolution decreases and more energy is reflected by the lens with the inevitable impedance mismatch.

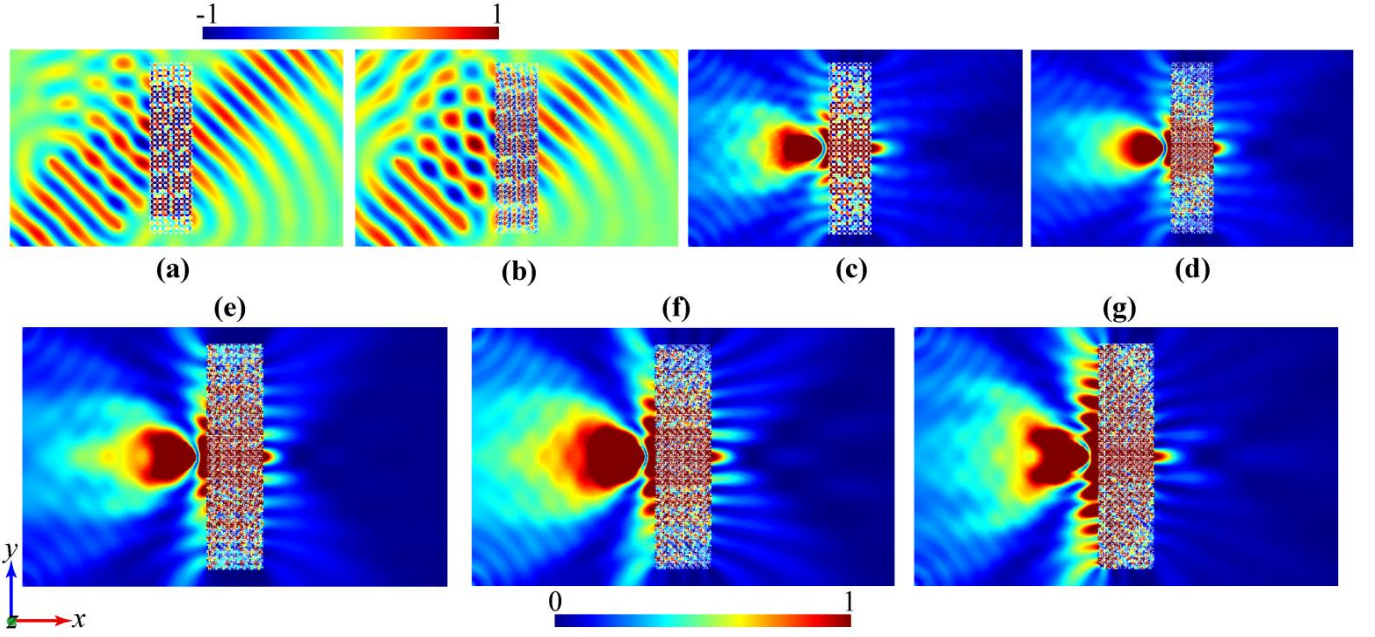

**Fig. 11. Simulations of negative refraction and acoustic subwavelength imaging using topology-optimized AMMs.** **(a)**-**(b)** Pressure fields under an incident Gaussian beam (45°) of acoustic waves for S1 at $\Omega$=0.173303 (a) and S12 at $\Omega$=0.160925 (b), respectively. **(c)** Imaging field pattern (the source is located at the position 2$a$ away from the left side of the 32×8 metamaterial slab) for S1 at $\Omega$=0.173303 (FWHM=0.44$\lambda$). **(d)-(g)** Imaging field patterns for S12 at $\Omega$=0.160925 (FWHM=0.39$\lambda$) (d), 0.173303 (FWHM=0.36$\lambda$) (e), 0.185682 (FWHM=0.42$\lambda$) (f) and 0.198061 (FWHM=0.44$\lambda$) (g), respectively.

To make clear the subdiffraction-limit resolution in Fig. 11, we present in Fig. 12 the zero-order transmission coefficient (Fang et al., 2006; Christensen et al., 2012; Shen et al., 2015) of a plane wave for evaluating the transmission of both propagating and evanescent waves through an 8-layered metamaterials of S1 or S12 immersed in air. Note that a value of zero-order transmission larger than 1.0 implies the enhancement of the propagating or evanescent waves. The regions representing the propagating waves locate in the left of the skew lines in Figs. 12(a) and 12(b). And the regions located in the right of the skew lines describe the case of evanescent waves. For every frequency within the double-negative range, it is clearly that the transmission coefficient can be larger than 1.0 for the wave vector $k_y$ either near or far away from $k_0$. And the lower-frequency range has the enhancements in the wider range of $k_y$ than the higher-frequency one. This emphasizes the importance of frequency in enhancing evanescent waves for imaging. Consequently, benefiting from the enhancement of evanescent waves, the metamaterial lens can capture the subwavelength information of the object and then transfer the corresponding energy to the focal plane of the image.

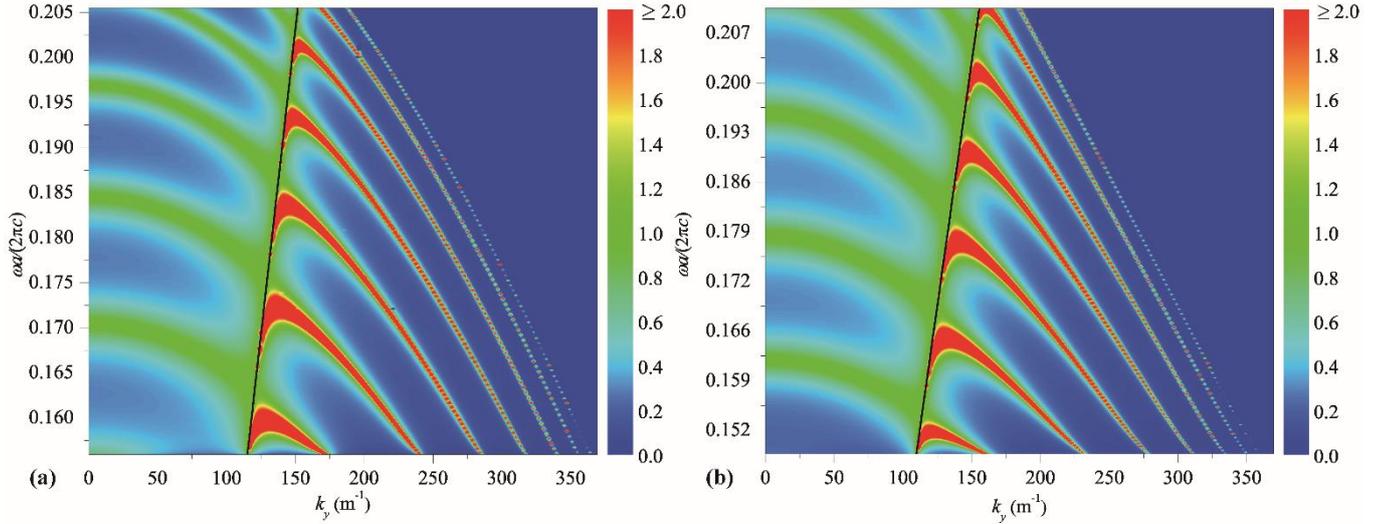

**Fig. 12. Enhanced transmission of the evanescent waves through topology-optimized AMMs.** Frequency and wave-vector dependence of the zero-order transmission coefficient for both propagating and evanescent waves through a layer consisting of 8 metamaterials S1 **(a)** or S12 **(b)**. Two skew lines in (a) and (b) represent the dispersion curves for air. If $k_y \leq k_0$ ($k_0$ is the propagation constant of the fundamental waveguide mode), the transmission coefficient characterizes the transmission property for the propagating waves, while for $k_y \leq k_0$ the corresponding waves represent the evanescent waves.

### 3.4.2. Experimental verification of acoustic subwavelength imaging

Above results show that the space-coiling AMMs can realize the ideal broadband double negativity with the relatively large air channels. In addition, they are mainly composed of lots of straight hard solid platess, showing the good workability. To show the convincing potential of the optimized AMMs, we worked with the space-coiling AMM S12 and experimentally demonstrated the broadband subwavelength imaging in Fig. 13. We adopted 3D-priting to fabricate the AMM sample made of polylactice acid (PLA) with mass density 1250 kg/m$^3$ and bulk modulus $3.5 \times 10^9$ Pa. The fabricated metamaterial slab in Figs. 13(a) and 13(c) consists of $20 \times 5$ microstructures depicted in Fig. 13(b). Experimental apparatus for the acoustic experiment inside a waveguide is illustrated in Fig. 13(a) where the slab sample was surrounded by the acoustic absorbing foams to avoid reflections in experiments. A loudspeaker located 3cm away from the input interface of the slab was used as the point source of waves; while the mounted microphone measured the acoustic field by moving in the scanning area. The measured result at each position was averaged over four measurements. Using the Fourier transform, the whole acoustic filed was obtained after the scanning measurement.

We first tested the subwavelength imaging of the metamaterial slab at 2200 Hz which is within the double-negative frequency range. The measured results in Fig. 13(e) agree well with the simulation results in Fig. 13(d) in terms of the acoustic magnitude field for the dashed area. The measured results at 2350 Hz in Fig. 13(f) also show the desired imaging pattern. The measured imaging resolutions of Figs. 13(e) and 13(f) are 0.38$\lambda$ and 0.44$\lambda$, respectively, certifying the subwavelength property well. Then we study the performance of the subwavelength imaging within the range of [1700Hz, 2500Hz], as displayed in Fig. 13(g). One peak with high transmission can be clearly observed at all measured frequencies, which demonstrates the broadband characteristic of the subwavelength imaging. Moreover, the non-monotonic curve in Fig. 13(h) exhibits all imaging resolutions less than or equal to the diffraction limit. Overall, the low operating frequency is beneficial to the high resolution. Clearly, the measured subwavelength imaging in Fig. 13 is owing to the double negativity of the metamaterials.

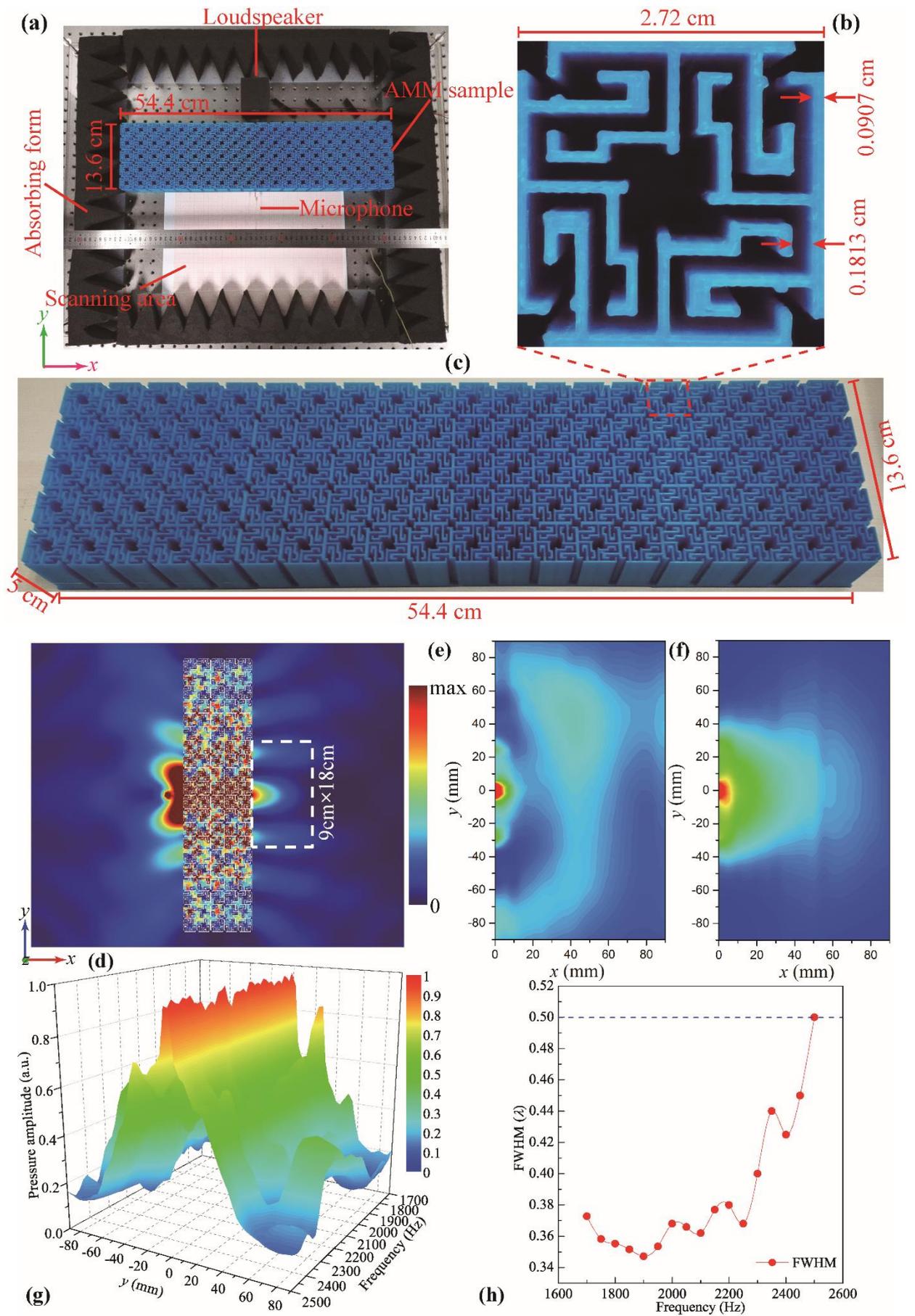

**Fig. 13. Experimental demonstration of acoustic subwavelength imaging using topology-optimized AMM S12. (a)** Experimental

apparatus. A loudspeaker acting as a point source is placed 3cm away from the metamaterial slab. **(b)** Fabricated 3D-printing microstructure of S12 using the polylactice acid (PLA). **(c)** Top view of the metamaterial slab. **(d)** Simulation result of 20×5 metamaterial slab based on S12 at 2200 Hz. **(e)-(f)** Measured magnitude fields at 2200 Hz (e) and 2350 Hz (f) of the dashed region in (d). Their corresponding imaging resolutions are FWHM=0.38$\lambda$ (e) and FWHM=0.44$\lambda$ (f), respectively. **(g)** Measured magnitude fields along the exiting surface of the metamaterial slab within the operating frequency range of [17000 Hz, 2500 Hz]. **(h)** Measured imaging resolutions within the operating frequency range.

## 4. Conclusions

In summary, for the first time, we construct a unified topology optimization framework for systematic designing the double negativity with any manual requirement including the expected microstructure symmetry, derivable double-negative mechanisms, necessary structural feature sizes and dispersion extent control of effective parameters. We design lots of novel microstructures with broadband double negativity and reveal the most beneficial topological features of resonance-cavity-based and space-coiling metamaterials. One feasible design principle is the suitable assembling of the multiple air resonant cavities, several solid blocks and air channels for resonance-cavity-based structures. Alternatively, one can realize the suitable combinations of the zigzag channels, thin curved hard solid plates and interconnection core regions in the centers. Exhaustive characterizations of the metamaterials indicate that the double negativity originating from the novel multipolar LC or Mie resonances can be induced by the simultaneous increasing or non-increasing mechanisms in optimization. Desired acoustic negative refraction and subwavelength imaging of optimized AMMs are numerically demonstrated in details for two representative AMMs. The enhancements of evanescent waves propagating through the metamaterials are found to be responsible for the subdiffraction-limit imaging resolution. More importantly, we also experimentally validate the broadband subwavelength imaging of the space-coiling AMMs.

It should also be pointed out that the developed topology optimization framework involving the derivable LC and Mie resonances is not restricted to the double negative metamaterials presented here. In principle, the design strategy proposed here can be universal for all types of AMMs demanding negative constitutive parameters, whether they are double-negative (Yang et al., 2013; Brunet et al., 2015), single-negative (Fang et al., 2006; Lu et al.; 2013; Shen et al., 2015) or even hyperbolic (Christensen et al., 2012; Shen et al., 2015; Christiansen et al., 2016). The present optimized AMMs and superlens provide the subwavelength imaging with the powerful and heuristic components, pushing the concept design to the specific practical applications. Our future work will focus on the in-depth design and realization of three-dimensional double-negative AMMs by topology optimization.

## Acknowledgements

This work is supported by the National Natural Science Foundation of China (Grant Nos. 11802012 and 11532001), Project funded by China Postdoctoral Science Foundation (2017M620607) and the Sino-German Joint Research Program (Grant No. 1355) and the German Research Foundation (DFG, Project No. ZH 15/27-1). H. W. Dong is also supported by the Fundamental Research Funds for the Central Universities (FRF-TP-17-070A1). H. W. Dong would like to thank Dr. Chen Shen (Duke University, USA) and Prof. Rui Zhu (Beijing Institute of Technology, P. R. China) for their helpful discussions.

## References

Aage, N., Andreassen, E., Lazarov, B. S., Sigmund, O., 2017. Giga-voxel computational morphogenesis for structural design. Nature 550, 84.

Brunet, T., Merlin, A., Mascaro, B., Zimny, K., Leng, J., Poncelet, O., Mondain-Monval, O., 2015. Soft 3D acoustic metamaterial with negative index. Nat. Mater. 14(4), 384.

Christensen, J., de Abajo, F.J.G., 2012. Anisotropic metamaterials for full control of acoustic waves. Phys. Rev. Lett. 108, 124301.

Cummer, S. A., Christensen, J., Alù, A., 2016. Controlling sound with acoustic metamaterials. Nat. Rev. Mater. 1, 16001.

Christiansen, R. E., Sigmund, O., 2016. Designing meta material slabs exhibiting negative refraction using topology optimization. Struct. Multidiscip. O. 54, 469-482.

Cheng, Y., Zhou, C., Yuan, B. G., Wu, D. J., Wei, Q., Liu, X. J., 2015. Ultra-sparse metasurface for high reflection of low-frequency sound based on artificial Mie resonances. Nat. Mater. 14, 1013.

Dong, H. W., Zhao, S. D., Wang, Y. S., Zhang, C., 2017. Topology optimization of anisotropic broadband double-negative elastic metamaterials. J. Mech. Phys. Solids 105, 54-80.

Dong, H. W., Zhao, S. D., Wang, Y. S., Zhang, C., 2018. Broadband single-phase hyperbolic elastic metamaterials for super-resolution imaging. Sci. Rep. 8, 2247.

Dong, H. W., Su, X. X., Wang, Y. S., Zhang, C., 2014a. Topological optimization of two-dimensional phononic crystals based on the finite element method and genetic algorithm. Struct. Multidiscip. O. 50, 593-604.

Dong, H. W., Su, X. X., Wang, Y. S., 2014b. Multi-objective optimization of two-dimensional porous phononic crystals. J. Phys. D Appl. Phys. 47(15), 155301.

Fang, N., Xi, D., Xu, J., Ambati, M., Srituravanich, W., Sun, C., Zhang, X., 2006. Ultrasonic metamaterials with negative modulus. Nat. Mater. 5, 452.

Frenzel, T., Kadic, M., Wegener, M., 2017. Three-dimensional mechanical metamaterials with a twist. Science, 358, 1072-1074.

Fokin, V., Ambati, M., Sun, C., Zhang, X., 2007. Method for retrieving effective properties of locally resonant acoustic metamaterials. Phys. Rev. B 76, 144302.

Goffaux, C., Vigneron, J. P., 2001. Theoretical study of a tunable phononic band gap system. Phys. Rev. B 64, 075118.

Graciá-Salgado, R., García-Chocano, V. M., Torrent, D., Sánchez-Dehesa, J., 2013. Negative mass density and ρ-near-zero quasi-two-dimensional metamaterials: Design and applications. Phys. Rev. B 88(22), 224305.

Guo, X., Zhang, W., Zhong, W., 2014. Doing topology optimization explicitly and geometrically—a new moving morphable components based framework. J. Appl. Mech. 81(8), 081009.

Han, T., Bai, X., Thong, J. T., Li, B., Qiu, C. W., 2014. Full control and manipulation of heat signatures: Cloaking, camouflage and thermal metamaterials. Adv. Mater. 26, 1731-1734.

Kumar, S., Bhushan, P., Prakash, O., Bhattacharya, S., 2018. Double negative acoustic metastructure for attenuation of acoustic emissions. Appl. Phys. Lett., 112, 101905.

Kaina, N., Lemoult, F., Fink, M., Lerosey, G., 2015. Negative refractive index and acoustic superlens from multiple scattering in single negative metamaterials. Nature, 525, 77.

Liang, Z., Li, J., 2012. Extreme acoustic metamaterial by coiling up space. Phys. Rev. Lett. 108, 114301.

Liu, T., Zhu, X., Chen, F., Liang, S., Zhu, J., 2018. Unidirectional Wave Vector Manipulation in Two-Dimensional Space with an All Passive Acoustic Parity-Time-Symmetric Metamaterials Crystal. Phys. Rev. Lett. 120, 124502.

Liu, Z., Zhang, X., Mao, Y., Zhu, Y. Y., Yang, Z., Chan, C. T., Sheng, P., 2000. Locally resonant sonic materials. Science 289, 1734-1736.

Lai, Y., Wu, Y., Sheng, P., Zhang, Z. Q., 2011. Hybrid elastic solids. Nat. Mater. 10, 620.

Lee, S. H., Choi, M., Kim, T. T., Lee, S., Liu, M., Yin, X., Zhang, X., 2012. Switching terahertz waves with gate-controlled active graphene metamaterials. Nat. Mater. 11, 936.

Li, Y., Yu, G., Liang, B., Zou, X., Li, G., Cheng, S., Cheng, J., 2014. Three-dimensional ultrathin planar lenses by acoustic metamaterials. Sci. Rep. 4, 6830.

Li, J., Fok, L., Yin, X., Bartal, G., Zhang, X., 2009. Experimental demonstration of an acoustic magnifying hyperlens. Nat. Mater. 8, 931.

Li, D., Zigoneanu, L., Popa, B. I., Cummer, S. A., 2012. Design of an acoustic metamaterial lens using genetic algorithms. J. Acoust. Soc. Am. 132, 2823-2833.

Li, Y.F., Huang, X., Meng, F., Zhou, S., 2016. Evolutionary topological design for phononic band gap crystals. Struct. Multidiscip. O. 54, 595-617.

Lee, S. H., Park, C. M., Seo, Y. M., Wang, Z. G., Kim, C. K., 2010. Composite acoustic medium with simultaneously negative density and modulus. Phys. Rev. Lett. 104, 054301.

Lanoy, M., Pierrat, R., Lemoult, F., Fink, M., Leroy, V., Tourin, A., 2015. Subwavelength focusing in bubbly media using broadband time reversal. Phys. Rev. B 91, 224202.

Lu, L., Yamamoto, T., Otomori, M., Yamada, T., Izui, K., Nishiwaki, S., 2013. Topology optimization of an acoustic metamaterial with negative bulk modulus using local resonance. Finite Elem. Anal. Des. 72, 1-12.

Matlack, K. H., Serra-Garcia, M., Palermo, A., Huber, S. D., Daraio, C., 2018. Designing perturbative metamaterials from discrete models. Nat. Mater. 17, 323-328.

Ma, G., Fan, X., Sheng, P., Fink, M., 2018. Shaping reverberating sound fields with an actively tunable metasurface. P. Natl. Acad. Sci., 201801175.

Mei, J., Ma, G., Yang, M., Yang, Z., Wen, W., Sheng, P., 2012. Dark acoustic metamaterials as super absorbers for low-frequency sound. Nat. Commun. 3, 756.

Molerón, M., Daraio, C., 2015. Acoustic metamaterial for subwavelength edge detection. Nat. Commun. 6, 8037.

Otomori, M., Yamada, T., Izui, K., Nishiwaki, S., Andkjær, J., 2012). A topology optimization method based on the level set method for the design of negative permeability dielectric metamaterials. Comput. Method. Appl. M. 237, 192-211.

Otomori, M., Yamada, T., Izui, K., Nishiwaki, S., Andkjær, J., 2017. Topology optimization of hyperbolic metamaterials for an optical hyperlens. Struct. Multidiscip. O. 55, 913-923.

Pendry, J.B., 2000. Negative refraction makes a perfect lens. Phys. Rev. Lett. 85, 3966.

Popa, B. I., Cummer, S. A., 2014. Non-reciprocal and highly nonlinear active acoustic metamaterials. Nat. Commun. 5, 3398.

Park, J. J., Park, C. M., Lee, K. J. B., Lee, S. H., 2015. Acoustic superlens using membrane-based metamaterials. Appl. Phys. Lett. 106, 051901.

Popa, B. I., Cummer, S. A., 2009. Design and characterization of broadband acoustic composite metamaterials. Phys. Rev. B 80, 174303.

Qi, S., Oudich, M., Li, Y., Assouar, B., 2016. Acoustic energy harvesting based on a planar acoustic metamaterial. Appl. Phys. Lett. 108, 263501.

Song, B. H., Bolton, J. S., 2000. A transfer-matrix approach for estimating the characteristic impedance and wave numbers of limp and rigid porous materials. J. Acoust. Soc. Am. 107, 1131-1152.

Shen, C., Xie, Y., Sui, N., Wang, W., Cummer, S. A., Jing, Y., 2015. Broadband acoustic hyperbolic metamaterial. Phys. Rev. Lett. 115, 254301.

Sanchis, L., García-Chocano, V. M., Llopis-Pontiveros, R., Climente, A., Martínez-Pastor, J., Cervera, F., Sánchez-Dehesa, J., 2013. Three-dimensional axisymmetric cloak based on the cancellation of acoustic scattering from a sphere. Phys. Rev. Lett. 110, 124301.

Shen, C., Xie, Y., Li, J., Cummer, S. A., Jing, Y., 2016. Asymmetric acoustic transmission through near-zero-index and gradient-index metasurfaces. Appl. Phys. Lett. 108, 223502.

Valentine, J., Zhang, S., Zentgraf, T., Ulin-Avila, E., Genov, D. A., Bartal, G., Zhang, X., 2008. Three-dimensional optical metamaterial with a negative refractive index. Nature 455, 376.

Wang, F., 2018. Systematic design of 3D auxetic lattice materials with programmable Poisson's ratio for finite strains. J. Mech. Phys. Solids 114, 303-318.

Xie, Y., Wang, W., Chen, H., Konneker, A., Popa, B. I., Cummer, S. A., 2014. Wavefront modulation and subwavelength diffractive acoustics with an acoustic metasurface. Nat. Commun. 5, 5553.

Xie, Y., Popa, B. I., Zigoneanu, L., Cummer, S. A., 2013. Measurement of a broadband negative index with space-coiling acoustic metamaterials. Phys. Rev. Lett. 110, 175501.

Yang, M., Ma, G., Yang, Z., Sheng, P., 2013. Coupled membranes with doubly negative mass density and bulk modulus. Phys. Rev. Lett. 110, 134301.

Yang, X., Kim, Y. Y., 2018. Topology optimization for the design of perfect mode-converting anisotropic elastic metamaterials. Compos. Struct. 201, 161-177.

Zhu, R., Liu, X. N., Hu, G. K., Sun, C. T., Huang, G. L., 2014. Negative refraction of elastic waves at the deep-subwavelength scale in a single-phase metamaterial. Nat. Commun. 5, 5510.

Zhang, X., Liu, Z., 2008. Superlenses to overcome the diffraction limit. Nat. Mater. 7(6), 435.

Zhang, S., Xia, C., Fang, N., 2011. Broadband acoustic cloak for ultrasound waves. Phys. Rev. Lett. 106, 024301.

Zigoneanu, L., Popa, B. I., Cummer, S. A., 2011. Design and measurements of a broadband two-dimensional acoustic lens. Phys. Rev. B 84, 024305.

Zhu, J., Christensen, J., Jung, J., Martin-Moreno, L., Yin, X., Fok, L., Garcia-Vidal, F. J., 2011. A holey-structured metamaterial for acoustic deep-subwavelength imaging. Nat. Phys. 7, 52.

Zhou, S., Li, W., Chen, Y., Sun, G., Li, Q., 2011. Topology optimization for negative permeability metamaterials using level-set algorithm. Acta. Mater. 59(7), 2624-2636.

Zhang, X., He, J., Takezawa, A., Kang, Z., 2018. Robust topology optimization of phononic crystals with random field uncertainty. Int. J. Numer. Methods Eng., 1-20.

**Appendix A. Performances of all optimized AMMs with double nativities**

| Number of structures | Topology of microstructures | Mechanism (Resonance) | Double-negative range in the target spectrum | Negative range beyond the target spectrum | Key optimization parameter |
|---|---|---|---|---|---|
| S1 | 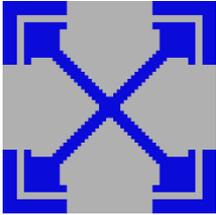 | LC<br>$\rho<0$: Dipolar<br>$K<0$: Quadrupolar | [0.156384, 0.198061] | [0.198061, 0.226167] | $\alpha = 1.0$<br>$w_a^* = a/30$<br>$\beta > 0$ |
| S2 | 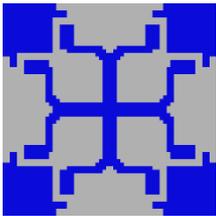 | LC<br>$\rho<0$: Dipolar<br>$K<0$: Quadrupolar | [0.177318, 0.198061] | [0.198061, 0.312607] | $\alpha = 1.5$<br>$w_a^* = a/30$<br>$\beta > 0$ |
| S3 | 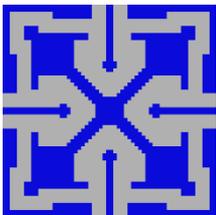 | LC<br>$\rho<0$: Dipolar<br>$K<0$: Quadrupolar | [0.190033, 0.198061] | [0.198061, 0.311203] | $\alpha = 1.0$<br>$w_a^* = a/15$<br>$\beta > 0$ |
| S4 | 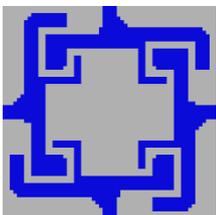 | LC<br>$\rho<0$: Quadrupolar<br>$K<0$: Monopolar+ Quadrupolar | [0.13852, 0.198061] | [0.198061, 0.208064] | $\alpha = 0.5$<br>$w_a^* = a/30$<br>$\beta > 0$ |
| S5 | 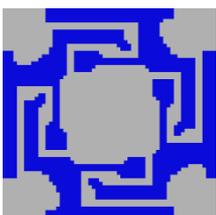 | LC<br>$\rho<0$: Quadrupolar<br>$K<0$: Monopolar+ Quadrupolar | [0.133584, 0.198061] | [0.198061, 0.199365] | $\alpha = 1.0$<br>$w_a^* = a/30$<br>$\beta > 0$ |
| S6 | 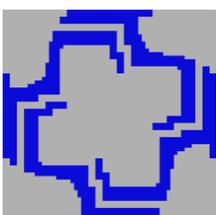 | LC<br>$\rho<0$: Quadrupolar<br>$K<0$: Monopolar+ Quadrupolar | [0.145441, 0.198061] | [0.198061, 0.210904] | $\alpha = 1.5$<br>$w_a^* = a/30$<br>$\beta > 0$ |
| S7 | 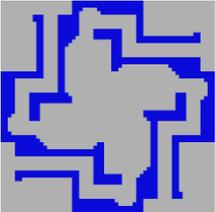 | LC<br>ρ<0: Quadrupolar<br>K<0: Monopolar+<br>Quadrupolar | [0.167848, 0.198061] | [0.198061, 0.264008] | $\alpha = 1.0$<br>$w_a^* = a/15$<br>$\beta > 0$ |
| S8 | 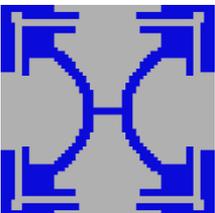 | LC<br>ρ<0: Dipolar<br>K<0: Quadrupolar | [0.148489, 0.198061] | [0.198061, 0.213618] | $\alpha = 1.0$<br>$w_a^* = a/30$<br>$\beta > 0$ |
| S9 | 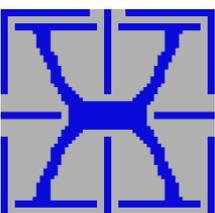 | LC<br>ρ<0: Dipolar<br>K<0: Quadrupolar | [0.140899, 0.198061] | [0.198061, 0.212211] | $\alpha = 1.5$<br>$w_a^* = a/30$<br>$\beta > 0$ |
| S10 | 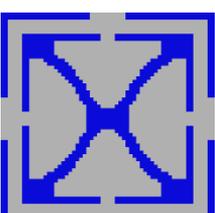 | LC<br>ρ<0: Dipolar<br>K<0: Quadrupolar | [0.176696, 0.198061] | [0.198061, 0.243152] | $\alpha = 1.0$<br>$w_a^* = a/15$<br>$\beta > 0$ |
| S11 | 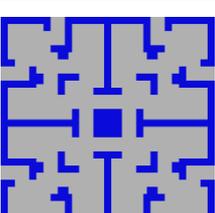 | Mie<br>ρ<0: Quadrupolar<br>K<0: Monopolar+<br>Quadrupolar | None | [0.25438, 0.379423] | $\alpha = 0$<br>$w_a^* = a/15$<br>$\beta \leq 0$ |
| S12 | 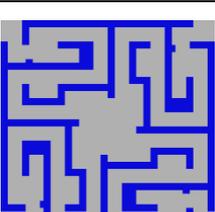 | Mie<br>ρ<0: Quadrupolar<br>K<0: Monopolar+<br>Quadrupolar | [0.147504, 0.198061] | [0.198061, 0.263384] | $\alpha = 0$<br>$w_a^* = a/15$<br>$\beta \leq 0$ |
| S13 | 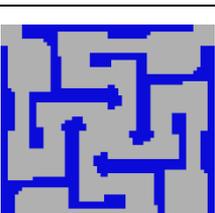 | Mie<br>ρ<0: Quadrupolar<br>K<0: Monopolar+<br>Quadrupolar | [0.191479, 0.198061] | [0.198061, 0.354621] | $\alpha = 1.0$<br>$w_a^* = a/15$<br>$\beta \leq 0$ |

| S14 | 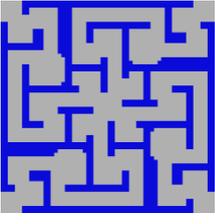 | Mie<br>ρ<0: Quadrupolar<br>$K$<0: Monopolar+<br>Quadrupolar | [0.156749, 0.198061] | [0.198061, 0.280516] | $\alpha = 0$<br>$w_a^* = a/15$<br>$\beta \leq 0$<br>$\Omega^* = 0.09903$ |
|---|---|---|---|---|---|
| S15 | 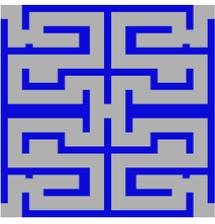 | Mie<br>ρ<0: Quadrupolar<br>$K$<0: Monopolar+<br>Quadrupolar | [0.146731, 0.198061] | [0.198061, 0.267489] | $\alpha = 0$<br>$w_a^* = a/15$<br>$\beta \leq 0$ |
| S16 | 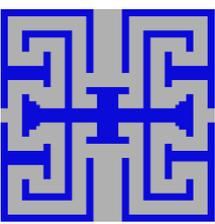 | Mie<br>ρ<0: Quadrupolar<br>$K$<0: Monopolar+<br>Quadrupolar | [0.161667, 0.198061] | [0.198061, 0.270861] | $\alpha = 1.0$<br>$w_a^* = a/15$<br>$\beta \leq 0$ |